\documentclass{article}%
\usepackage{graphicx}
\usepackage{amsfonts}
\usepackage{geometry}
\usepackage{color}
\usepackage{amsmath}
\usepackage{authblk}

\usepackage{verbatim}
\usepackage{amssymb}%
\providecommand{\U}[1]{\protect\rule{.1in}{.1in}}
\begin{document}
\title{Constraint algebra in LQG reloaded : Toy model of an Abelian gauge theory - II\\Spatial Diffeomorphisms}

\author[a]{Adam Henderson}
\author[b,a]{Alok Laddha}
\author[a,c]{Casey Tomlin}
\affil[a]{Institute for Gravitation and the Cosmos\\Pennsylvania State University, University Park, PA 16802-6300, U.S.A}
\affil[b]{Chennai Mathematical Institute\\ Siruseri, Chennai-603103, India}
\affil[c]{Max Planck Institute for Gravitational Physics (Albert Einstein Institute) Am M\"{u}Ÿhlenberg 1, D-14476 Potsdam, Germany}

\maketitle
\begin{abstract}
In \cite{hat} we initiated an approach towards quantizing the Hamiltonian constraint in Loop Quantum Gravity (LQG) by requiring  that it generates an anomaly-free representation of constraint algebra off-shell. We investigated this issue in the case of a toy model of a $2+1$-dimensional $U(1)^{3}$ gauge theory, which can be thought of as a weak coupling limit of Euclidean three dimensional gravity. However in \cite{hat} we only focused on the most non-trivial part of the constraint algebra that involves commutator of two Hamiltonian constraints.\\
In this paper we continue with our analysis and obtain a representation of  full constraint algebra in loop quantized framework. We show that there is a representation of the Diffeomorphism group with respect to which the Hamiltonian constraint quantized in \cite{hat} is diffeomorphism covariant.
Our work can be thought of as a potential first step towards resolving some long standing issues with the Hamiltonian constraint in canonical LQG.
\end{abstract}

\section{Introduction}

A satisfactory definition of Hamiltonian constraint in Loop Quantum Gravity (LQG) \cite{ttbook} remains an open problem. Despite remarkable progress made in the seminal work of Thiemann (\cite{qsd1},\cite{qsd2},\cite{qsd3}), 
it is clear that the current quantization is not satisfactory due to three inter related issues :  \noindent {(1)} Enormous ambiguity in the definition of the continuum Hamiltonian constraint, \noindent {(2)} The absence of a representation of Quantum Dirac algebra (referred to as the off-shell closure in \cite{nicolaietal}), and \noindent {(3)} When the constraint is used in symmetry reduced sector of Loop Quantum cosmology, the low energy limit of the theory turns out to be incorrect, \cite{improvedlqc}.
Progress in obtaining a satisfactory definition of quantum dynamics in canonical LQG can be achieved by analyzing and overcoming the first two
obstacles  by taking hints from  toy models like Loop Quantum Cosmology.\\
In \cite{hat},\cite{madtom} a new approach was initiated to quantize the Hamiltonian constraint in LQG. This approach is based on the lessons learnt in (\cite{pftham}, \cite{ttpft}, \cite{amdiff}, \cite{improvedlqc}). The idea  in \cite{hat} was to look for higher density constraints whose action at finite triangulation was based upon the geometric action of the classical constraints on phase space fields. The continuum limit of finite triangulation constraint is taken \emph{not on} ${\cal H}_{kin}$ but on certain distributional subspaces known as habitats \cite{lm}. Instead of working with full LQG,in \cite{hat} we considered a simple toy model of 2+1 dimensional\footnote{As we insisted in \cite{hat} and would like to remind the reader  again here that although the model we consider is 2+1 dimensional theory, our analysis to a large extent is independent of dimensionality and we believe it goes through rather straight-forwardly in 3+1 dimensions. Infact as we argued in \cite{hat}, some of the technicalities which are present in two spatial dimensions will be absent in three dimensions, thus simplifying the analysis. This should not be too surprising as off-shell closure of Dirac algebra probes the local structure of field theory, even when the theory is topological \emph{on-shell}} $U(1)^{3}$ gauge theory which can be thought of as a weak coupling limit of Euclidean canonical gravity. \cite{smolin-weakcouple}\\
In \cite{hat} we showed that there exists  quantization of  (density weight $\frac{5}{4}$) Hamiltonian constraint which satisfied,
\begin{equation}
[\hat{H}[N], \hat{H}[M]]\ =\ \widehat{H_{diff}[q^{-1}(N\nabla M\ -\ M\nabla N)]}
\end{equation}
In this paper we continue the analysis of obtaining a representation of the
constraint algebra in the loop quantized U$(1)^{3}$ gauge theory. Our goal is to obtain a
representation of the \textquotedblleft Dirac algebra\textquotedblright in the following sense.\footnote{The quotation marks indicate that strictly speaking we are not working with the algebra of constraints but with the crossed product generated by the Hamiltonian constraint and finite diffeomorphisms.} %
\begin{align}
\hat{U}(\phi_{1})\hat{U}(\phi_{2})  &  =\hat{U}(\phi_{1}\circ\phi_{2})\\
\hat{U}(\phi)^{-1}\hat{H}[N]\hat{U}(\phi)  &  =\hat{H}[\phi^{\ast
}N]\label{diracalg2}\\
\lbrack\hat{H}[N],\hat{H}[M]]  &  =\mathrm{i}\hbar\hat{D}[\hat{\vec{\omega}}]
\label{diracalg3}%
\end{align}
In this paper we focus on the (spatial) diffeomorphism covariance of
the Hamiltonian constraint. That is, we want to see if there exists a
representation of Diff$(\Sigma)$ on $\mathcal{V}_{\mathrm{LMI}}$ such that
(\ref{diracalg2}) is represented without anomaly.\\

Right at the outset, it appears that the answer will be in the negative, due
to background structure which
is required to define the quantum Hamiltonian constraint in \cite{hat}.

\begin{enumerate}
\item[(1)] \emph{Diffeomorphism non-covariance of the quantum shift}: The
action of the (finite-triangulation) Hamiltonian constraint on a charge
network $c$ results in a deformation of the underlying graph $\gamma(c)$ in a
neighborhood of vertices of $\gamma(c)$ in the direction of vectors which are
themselves defined using data from $c$. Given a charge network state
$|c\rangle$ and its vertex set $V(c)$, the (regularized) expectation value
$\langle c|\hat{E}_{i}^{a}\hat{q}^{-1/4}(v)|_{\epsilon}|c\rangle$ at any
vertex $v\in V(c)$ plays the role of this vector, and was referred to as the
quantum shift in \cite{hat}. The subscript $\epsilon$ indicates that the
operator implicitly depends on regulating structures which are parametrized by
$\epsilon$. As we show below, due to the regularization dependence of the quantum shift vector, it turns out that,
given a state $|\phi\cdot c\rangle$ (the diffeomorphic image of $c$ under
$\phi$), the quantum shift $\langle\phi\cdot c|\hat{E}_{i}^{a}\hat{q}%
^{-1/4}(\phi v)|_{\epsilon}|\phi\cdot c\rangle$ defined at $\phi(v)\in
V(\phi\cdot c)$ is not the pushforward (via $\phi$) of $\langle c|\hat{E}%
_{i}^{a}\hat{q}^{-1/4}(v)|_{\epsilon}|c\rangle$. This is the first obstruction
which, unless addressed, will ensure that the Hamiltonian constraint operator
will not commute with finite diffeomorphisms.\newline
\end{enumerate}

\emph{Non-covariant nature of extraordinary (EO) vertices}: The key feature of
the Hamiltonian constraint's action on kinematical states\footnote{By this we
mean the Hamiltonian constraint operator at finite triangulation which is
densely defined on $\mathcal{H}_{\mathrm{kin}}.$} is the creation of so-called
extraordinary (EO) vertices. Essentially the idea is the following: Starting
with any charge network $c,$ around each of its non-degenerate vertices
(vertices at which not all edges emanating from it are charged only in one
copy of U(1)), we fix, once and for all, a coordinate ball $B(v,\epsilon)$ of
radius $\epsilon$ and consider a sequence of finite-triangulation Hamiltonian
constraint operators $\hat{H}_{T(\delta)}[N]$ with $\delta\leq\delta
_{0}(\epsilon)$. The action of $\hat{H}_{T(\delta)}[N]$ on $|c\rangle$ creates
a linear combination of charge network states, each of which has an EO vertex
sitting inside $B(v,\epsilon)$. These EO vertices have several distinguishing properties:

\begin{enumerate}
\item[(1)] They are necessarily zero-volume.

\item[(2)] By construction they are inside $B(v,\epsilon)$ with $v$ being some
non-degenerate and have \textquotedblleft non-zero volume\footnote{In the
$\mathrm{U}(1)^{3}$ theory in 2+1 dimensions, the notion of degenerate vertex
and zero-volume vertex are not equivalent. As we have defined above, a
degenerate vertex is the one on which all incident edges are charged only in
one copy of $\mathrm{U}(1)$. All such degenerate vertices are necessarily
zero-volume; however one could easily have a zero volume vertex which was not
degenerate. The Hamiltonian constraint action on a charge network generically
created vertices which are degenerate, but in some special cases it creates
vertices which are zero volume but non-degenerate. We labelled them type-B EO
vertices in \cite{hat}.}\textquotedblright\ vertex.

\item[(3)] Their \textquotedblleft location\textquotedblright\ (with respect
to the fixed coordinate chart around $v$) is state-dependent and dictated by
the so-called quantum shift.

\item[(4)] Given any charge network $c^{\prime}$ with a vertex $v^{\mathrm{E}%
}$ satisfying the above three properties, one can always find charge network
$c$ with a vertex $v$ such that the action of $\hat{H}_{T(\delta)}(v)$ results
in a linear combination of charge network states including $|c^{\prime}%
\rangle$.
\end{enumerate}

Although the EO vertices are zero-volume, the action of the Hamiltonian
constraint on such vertices is required to be non-trivial and have a specific
form in order to obtain an anomaly-free commutator of two continuum
Hamiltonian constraints (for more details we urge the reader to consult
\cite{hat}). The definition of an EO vertex relies on the vertices lying
inside certain prescribed coordinate neighborhoods of non-degenerate vertices.
Whence under an arbitrary diffeomorphism, an EO vertex could be dragged
outside the the prescribed neighborhood and would no longer be classified as
an EO vertex by our prescription. This means that the action of a Hamiltonian
constraint on a state will not commute with the action of
diffeomorphisms. 

\emph{Non-trivial density weight of the Lapse} : As the Hamiltonian constraint $H(x)$ is a scalar density of weight $\frac{5}{4}$, the lapse function is a scalar density of weight $-\frac{1}{4}$ whence evaluation of Lapse at a given point requires an explicit specification of co-ordinate chart in the neighborhood of the point. Whence the pullback of a Lapse by a diffeomorphism which occurs in $\hat{H}[\phi^{*}N]$ involves Jacobian between various co-ordinate charts and it is not clear how such factors could arise in $\hat{U}(\phi)\hat{H}[N]\hat{U}(\phi^{-1})$. This is yet another potential source of diffeomorphism non-covariance of the Hamiltonian constraint.\\

In this paper we show that, despite the apparentbackground-dependence of the
quantum Hamiltonian constraint, we obtain an anomaly-free representation of
the Dirac algebra on $\mathcal{V}_{\mathrm{LMI}}$ by defining a new
representation of the diffeomorphism group on $\mathcal{H}_{\mathrm{kin}}$
(and whence by dual action on $\mathcal{V}_{\mathrm{LMI}}$).\\

Purpose of requiring the Hamiltonian constraint to be diffeomorphism constraint is two fold \cite{qsd1}. On the one hand, this ensures that the quantum constraint algebra is first class, and perhaps equally importantly, the vast amount of ambiguity which persists in the continuum quantum constraint can be reduced by requiring diffeomorphism covariance. This has been explicitly demonstrated in \cite{qsd1} and as we see below, it remains true even in our approach.\\

This paper is organized as follows: In Section \ref{RECAP} we recall key ideas
and results from \cite{hat}. That section merely serves to summarize contents
of \cite{hat} and we do not claim it to be a sufficient prerequisite for
understanding all the details in the subsequent sections. We urge the
interested reader to consult \cite{hat} for a more detailed understanding of
the structures involved. In Section \ref{NA} we do a sample
computation where we take the standard representation of diffeomorphism group
on $\mathcal{H}_{\mathrm{kin}}$ and obtain, via dual action, a conjugate
representation on $\mathcal{V}_{\mathrm{LMI}}$. We then check if the
Hamiltonian constraint operator $\hat{H}[N]$ is covariant under this
representation of the diffeomorphism group. As expected, $\hat{H}[N]$ is not covariant and
the analysis reveals precisely where the issues mentioned above show up in the
computation. In Section \ref{newrep}, we define a new representation of
$\mathrm{Diff}(\Sigma)$ on $\mathcal{V}_{\mathrm{LMI}}$ which essentially
ensures that various structures required to define the Hamiltonian constraint
operator at finite triangulation behave covariantly with respect to this
representation. In Section \ref{covariance}, which is the main section of the
paper, we prove the diffeomorphism-covariance of the continuum Hamiltonian
constraint on $\mathcal{V}_{\mathrm{LMI}}$. Together with \cite{hat}, the
results of this section establish a representation of the Dirac algebra for
the $U(1)^{3}$ gauge theory. In section \ref{spectrum} we perform a heuristic check on the validity of the new representation of $Diff(\Sigma)$ on ${\cal V}_{LMI}$ by computing a subset of physical states that capture the topological sector in the quantum theory and argue that the final answer we obtain is the expected one in the sense that we would have arrived at the same answer had we worked with the representation of $Diff(\Sigma)$ commonly used in LQG.\\
We end with conclusions where we highlight the unsatisfactory aspects of our work, which is the use of several auxiliary structures and make some remarks pertaining to the generalization of our work for Euclidean LQG.





\section{Summary of \cite{hat} and some Notational Changes}

\label{RECAP}

In this section, we briefly recap the relevant results and notation of
\cite{hat}, where more details can be found when desired. There, a proposal
was made for a finite-triangulation Hamiltonian constraint operator $\hat
{H}_{T(\delta)}$ on the vector space $\mathcal{D}$ spanned by charge networks
$c\equiv(c_{1},c_{2},c_{3})$ (whose completion is the kinematical Hilbert
space $\mathcal{H}_{\mathrm{kin}}$ of the theory) and its continuum limit
($\delta\rightarrow0$) was evaluated on a vector space $\mathcal{V}%
_{\mathrm{LMI}}\subset\mathcal{D}^{\ast}$ of distributions spanned by objects
of the type%
\begin{equation}
(\Psi_{\lbrack c]_{(1)}}^{f^{(1)}}|~:=\sum_{(c_{1}^{\prime},c_{2},c_{3}%
)\in\lbrack c]_{(1)}}f^{(1)}(\bar{V}(c_{1}^{\prime}\cup c_{2}\cup
c_{3}))\langle c_{1}^{\prime},c_{2},c_{3}|
\end{equation}
Here $f^{(1)}:\Sigma^{|V(c)|}\rightarrow%
\mathbb{C}
$ is a smooth function that is symmetric in it's arguments\footnote{This assumption was not made in \cite{hat} however it is invoked here in the interest of pedagogy. The analysis given in the paper can be easily seen to hold when we relax this assumption.}, and the set of arguments $\bar{V}(c_{1}^{\prime}\cup
c_{2}\cup c_{3})$ is a set of vertices (with the same cardinality as the
vertex set $V(c)$ of $c$) in which any WEO pairs (recalled below) of vertices
are replaced by the single WEO vertex of the pair. In this work we will omit
the parentheses around the U(1)$_{i}$ labels to slightly simplify the
notation; i.e., these states will be written $\Psi_{\lbrack c]_{i}}^{f^{i}}.$

$[c]_{(i=1)}$ is a set of charge networks containing a `parent' $c=(c_{1}%
,c_{2},c_{3}),$ as well all other charge networks $(c_{1}^{\prime},c_{2}%
,c_{3})$ in which $c_{2},c_{3}$ are unaltered, but $c_{1}^{\prime}\neq c_{1},$
and the non-equality is of a special type; namely, each $(c_{1}^{\prime}%
,c_{2},c_{3})$ has at least one vertex which is `weakly extraordinary (WEO) of
type $i=1$' with respect to $c.$ We digress briefly to explain the notion of
extraordinary and weakly extraordinary vertices.

Roughly speaking, extraordinary (EO) vertices of a charge network $c^{\prime}$
are those produced by the action of a finite-triangulation Hamiltonian
constraint operator on a charge network $c$, and WEO vertices are EO vertices
which have additionally been moved by diffeomorphisms which are the identity
on $c.$ Given a charge network and the associated coordinate charts based at
its vertices, there is a list of criteria (found in \cite{hat}) which
determines whether a vertex is WEO, EO, or neither. This is largely a
topological and charge label-dependent classification, and given an arbitrary
charge network $c$ with WEO vertices, it is possible to reconstruct a unique  WEO-vertex free 
charge network $\tilde{c}$. This comes about as follows. As shown in appendix B of \cite{hat}, any WEO vertex $v^{E}$ in a charge-network $c$ is uniquely associated to a vertex $v\ \in V( c)$. Furthermore all the (maximal analytic extension of) edges beginning at $v^{E}$ terminate in a three-valent vertex. By erasing each of this edge and adding the corresponding charge to the edge between the above mentioned three-valent vertex and $v$ one reconstructs a WEO-vertex free charge network that we will denote by $\tilde{c}$ throughout this paper.\\
If the classification scheme determines that a given vertex $v^{\mathrm{E}}$
is EO, then it is uniquely associated to another vertex $v$, namely that
vertex at which a finite-triangulation Hamiltonian-type operator has acted to
produce the pair $\left(  v,v^{\mathrm{E}}\right)  $. It is helpful to keep in
mind a picture of the action of an operator of the type $\hat{E}_{i}^{a}%
\hat{F}_{ab}^{j}\hat{E}_{k}^{b}$ for some fixed $i\neq j\neq k$ at a vertex
$v$ of some charge network $c$. In \cite{hat} we have constructed such
operators via a `loop assignment scheme' where, roughly speaking, $\hat{E}%
_{i}^{a}$ gives the direction and magnitude of one leg of the loops, $\hat
{F}_{ab}^{j}$ determines that the charge on the attached loops is only
non-zero in U(1)$_{j},$ and $\hat{E}_{k}^{b}$ determines the magnitude of
those U(1)$_{j}$ charges via the U(1)$_{k}$ charges on edges of the underlying
state. EO vertices are common apex points of the (charged) loop collections,
and hence come in several flavors, and it is necessary in what follows to
respect their distinction. To this end, we introduce some additional notation.

Each $v^{\mathrm{E}}$ is first classified as type A or type B; type A EO
vertices lie off of the original graph $c,$ and type B vertices lie on $c.$
This distinction is not important below, so we omit this information from our
notation, and focus the discussion on type A EO vertices $v^{\mathrm{E}}$. Let
$v_{\delta}^{\mathrm{E}}(j,k)$ denote an EO vertex from which all outgoing
edges are charged in U(1)$_{j},$\footnote{In \cite{hat}, we denoted this vertex as $v_{\delta}^{\mathrm{E}}(M,k)$.} with the magnitude of those charges being
determined by the U(1)$_{k}$ charges in the underlying charge network. The
subscript $\delta$ denotes that $v_{\delta}^{\mathrm{E}}(j,k)$ is located a
coordinate distance $\delta|\langle\hat{E}_{i}^{a}\rangle|$ (with respect to
the coordinate system based at the vertex $v$ of the underlying charge
network) from $v$ in the direction of $\langle\hat{E}_{i}^{a}\rangle.$ By the
classification scheme detailed in \cite{hat}, the pair $\left(  v,v_{\delta
}^{\mathrm{E}}(j,k)\right)  $ is unique, and we term it an extraordinary pair.

Weakly extraordinary (WEO) vertices are again detectable via the
classification scheme, and are generated from EO vertices by applying
diffeomorphisms (which do not move the underlying state) to states containing
EO pairs. That is, these diffeomorphisms only move the loop collections
produced by Hamiltonian-type actions.

In \cite{hat}, the calculation of the continuum limit action $\hat{H}%
\Psi_{\lbrack c]_{i}}$ (in this work we drop the prime on $\hat{H}$ as an
operator on $\mathcal{V}_{\mathrm{LMI}}$) was performed by first writing the
Hamiltonian as a sum $H=\sum_{i}H^{i}$, where $i$ labels the U(1) index
appearing on the curvature $F_{ab}^{i},$ and considering the various actions
$\hat{H}^{i}\Psi_{\lbrack c]_{j}}$ for each $i,j.$ The result is as follows:
Given a density-weight $-\frac{1}{2}$ lapse function $N,$ it was found that%
\begin{equation}
\hat{H}^{1}[N]\Psi_{\lbrack c]_{1}}^{f^{1}}=\sum_{v\in V(c)}\left(
\Psi_{\lbrack c]_{1}}^{f_{v,2}^{1,1}[N]}-\Psi_{\lbrack c]_{1}}^{f_{v,3}%
^{1,1}[N]}\right)  \label{H1onLMI}%
\end{equation}
where $f_{v,2}^{1,1}[N]$ is a (generally discontinuous) function which agrees
with $f^{1}$ at almost all points of $\Sigma^{|V(c)|},$ except when its
argument is $\bar{V}(c_{1}^{\prime}\cup c_{2}\cup c_{3})$ for some $(c_{1}^{\prime}%
,c_{2},c_{3})\in\lbrack c]_{1}$, and $v\in\bar{V}(c_{1}^{\prime}\cup c_{2}\cup
c_{3}),$ in which case it takes the value%
\begin{equation}
f_{v,2}^{1,1}[N]\left(\bar{V}(c_{1}^{\prime}\cup c_{2}\cup
c_{3})\right)  =N(v)\lambda(\vec
{n}_{v}^{c})\sum_{e\vert b(e)\ =\ v}n_{e}^{2}\dot{e}^{a}(0)\frac{\partial}{\partial
v^{a}}f^{1}\left(  v,\bar{V}(c_{1}^{\prime}\cup c_{2}\cup
c_{3})-\{v\}\right)  \label{functionforH1}%
\end{equation}
where $\dot{e}(0)$ is the vector tangent to edge $e$ at $v=b(e)$, and is assumed to be of unit length in a prescribed co-ordinate system.\\
In \cite{hat} this function was called $\bar{f}_{v}^{(1)(1)},$ where the
second superscripted $(1)$ refers to the action of $\hat{H}^{1},$ and the bar
to the fact that the directional derivative is along $\langle\hat{E}_{2}%
^{a}(v)\rangle_{c}$ (as opposed to along $\langle\hat{E}_{3}^{a}(v)\rangle
_{c},$ which in \cite{hat} was denoted by a double bar). The expressions for
$\hat{H}^{2}\Psi_{\lbrack c]_{2}}$ and $\hat{H}^{3}\Psi_{\lbrack c]_{3}}$ can
be obtained by cyclic permutation of the indices in the above equations.

The action of the mixed-index cases were found to be of the form%
\begin{equation}
\hat{H}^{2}[N]\Psi_{\lbrack c]_{1}}^{f^{1}}=\sum_{v\in V(c)}\left(
\Psi_{\lbrack c]_{1}}^{f_{v,1}^{1,2}[N]}-\Psi_{\lbrack c]_{1}}^{f_{v,3}%
^{1,2}[N]}\right)  \label{h2psi1}%
\end{equation}
where again, the functions $f_{v,1}^{1,2}[N],f_{v,3}^{1,2}[N]$ agree with
$f^{1}$ at all values of $\Sigma^{|V(c)|},$ except when those arguments
coincide with $\bar{V}(c_{1}^{\prime}\cup c_{2}\cup c_{3})$ with $v\in V(c),$
\emph{and} there is an EO vertex $v_{\delta}^{\mathrm{E}}(1,2)\in
\mathrm{supp}(N)$ associated with $v,$ in which case we have%
\begin{subequations}
\begin{align}
f_{v,1}^{1,2}[N]\left(  \bar{V}(c_{1}^{\prime}\cup c_{2}\cup c_{3})\right)
&  =\left[  \sum_{e\in E(c)|b(e)=v}n_{e}^{1}\dot
{e}^{a}(0)\partial_{a}N(v)\right]  f^{1}\left(  \bar{V}(c_{1}^{\prime}\cup
c_{2}\cup c_{3})\right) \label{h2psi1f1}\\
f_{v,3}^{1,2}[N]\left(  \bar{V}(c_{1}^{\prime}\cup c_{2}\cup c_{3})\right)
&  =\left[  \sum_{e\in E(c)|b(e)=v}n_{e}^{3}\dot
{e}^{a}(0)\partial_{a}N(v)\right]  f^{1}\left(  \bar{V}(c_{1}^{\prime}\cup
c_{2}\cup c_{3})\right)  \label{h2psi1f3}%
\end{align}
Recall that $n_{e}^{i}$ are charges on the edge $e\ \in\ E( c )$ in $U(1)_{i}$.
Similarly,%
\end{subequations}
\begin{equation}
\hat{H}^{3}[N]\Psi_{\lbrack c]_{1}}^{f^{1}}=\sum_{v\in V(c)}\left(
\Psi_{\lbrack c]_{1}}^{f_{v,2}^{1,3}[N]}-\Psi_{\lbrack c]_{1}}^{f_{v,1}%
^{1,3}[N]}\right)  \label{h3psi1}%
\end{equation}
with (under analogous conditions as stated above, with $v_{\delta}%
^{\mathrm{E}}(1,2)$ replaced with $v_{\delta}^{\mathrm{E}}(1,3)$)
\begin{subequations}
\begin{align}
f_{v,2}^{1,3}[N]\left(  \bar{V}(c_{1}^{\prime}\cup c_{2}\cup c_{3})\right)
&  =\left[  \sum_{e\in E(c)|b(e)=v}n_{e}^{2}\dot
{e}^{a}(0)\partial_{a}N(v)\right]  f^{1}\left(  \bar{V}(c_{1}^{\prime}\cup
c_{2}\cup c_{3})\right) \label{h3psi1f2}\\
f_{v,1}^{1,3}[N]\left(  \bar{V}(c_{1}^{\prime}\cup c_{2}\cup c_{3})\right)
&  =\left[  \sum_{e\in E(c)|b(e)=v}n_{e}^{1}\dot
{e}^{a}(0)\partial_{a}N(v)\right]  f^{1}\left(  \bar{V}(c_{1}^{\prime}\cup
c_{2}\cup c_{3})\right)  \label{h3psi1f1}%
\end{align}
The expressions for the remaining $\hat{H}^{i}[N]\Psi_{\lbrack c]_{j}}^{f^{j}%
}$ are cyclic permutations of these. Given these preliminaries, we now embark
on a first attempt (and failure) to arrive at a statement of diffeomorphism
covariance of this Hamiltonian.

\section{Naive Attempt}\label{NA}

We now quantify the worries laid out in the introduction regarding why, using
the usual representation of the group of semi-analytic diffeomorphisms (denoted in this paper by $\mathrm{Diff}(\Sigma)$) that is used in loop
quantum gravity, the Hamiltonian constraint constructed in \cite{hat} is not
diffeomorphism-covariant. More in detail, in this section we ask the following
question. Consider a representation of Diff$(\Sigma)$ on $\mathcal{V}%
_{\mathrm{LMI}}$ induced via dual action:%
\end{subequations}
\begin{equation}
\left(  \hat{U}(\phi)^{\prime}\Psi_{\lbrack\tilde{c}]_{i}}^{f^{i}}\right)
(|c\rangle):=\Psi_{\lbrack\tilde{c}]_{i}}^{f^{i}}(\hat{U}(\phi)|c\rangle)
\end{equation}
where the right hand side of the above equation is given by using the
\textquotedblleft natural\textquotedblright\ unitary representation of
Diff$(\Sigma)$ on $\mathcal{H}_{\mathrm{kin}}$ \cite{almmt}. We now ask if
$\hat{U}(\phi)^{\prime}\hat{H}^{j}[N]\hat{U}(\phi)^{-1\prime}\Psi
_{\lbrack\tilde{c}]_{i}}^{f^{i}}$ equals $\hat{H}^{j}[\phi^{\ast}%
N]\Psi_{\lbrack\tilde{c}]_{i}}^{f^{i}}$ for all $i,\ j$. As we will see, the
answer is no, and the reasons are precisely those which were given in the introduction.

Readers who are convinced by the arguments given in the introduction can
safely skip this section. However those who wish to follow details in the
subsequent sections might find it helpful to peruse the computations done
here. With out loss of generality, we restrict attention to $i=1$ and check
the diffeomorphism covariance of $\hat{H}^{j=1,2}[N]$ in the domain defined by
$\Psi_{\lbrack\tilde{c}]_{1}}^{f^{1}}$.

\subsection{Checking Diffeomorphism Covariance of $\hat{H}^{1}[N]$ on
$\Psi_{\lbrack\tilde{c}]_{1}}^{f^{1}}$}

\label{NA-1}

Given any charge network state $|c_{A}\rangle$ we would like to see if
\begin{equation}
\left(  \hat{U}(\phi)^{\prime}\hat{H}^{1}[N]\hat{U}(\phi)^{-1\prime}%
\Psi_{\lbrack\tilde{c}]_{1}}^{f^{1}}\right)  (|c_{A}\rangle)=\left(  \hat
{H}^{1}[\phi^{\ast}N]\Psi_{\lbrack\tilde{c}]_{1}}^{f^{1}}\right)
(|c_{A}\rangle) \label{naiveattempt1}%
\end{equation}
where $\hat{U}(\phi)^{\prime}$ denotes the natural representation of
Diff$(\Sigma)$ on $\mathcal{V}_{\mathrm{LMI}}$ obtained by dualizing the
action of Diff$(\Sigma)$ on $\mathcal{H}_{\mathrm{kin}}$. We can deduce this
representation as follows:%
\begin{align}
\left(  \hat{U}(\phi)^{\prime}\Psi_{\lbrack\tilde{c}]_{1}}^{f^{1}}\right)
(|c^{\prime}\rangle):=  &  \Psi_{\lbrack\tilde{c}]_{1}}^{f^{1}}(|\phi\cdot
c^{\prime}\rangle)\label{naturalrepdiff1}\\
=  &  \sum_{c^{\prime\prime}\in\lbrack\tilde{c}]_{1}}f^{1}(\bar{V}%
(c^{\prime\prime}))\delta_{c^{\prime\prime},\phi\cdot c^{\prime}}\nonumber\\
=  &  \left(  \Psi_{\phi^{-1}\cdot\lbrack\tilde{c}]_{1}}^{f^{1}\circ\phi
}\right)  (|c^{\prime}\rangle),\nonumber
\end{align}
where
\begin{equation}
\phi\cdot\lbrack\tilde{c}]_{1}=\{\phi\cdot(c_{1}^{\prime},\tilde{c}_{2}%
,\tilde{c}_{3})|(c_{1}^{\prime},\tilde{c}_{2},\tilde{c}_{3})\in\lbrack
\tilde{c}]_{1}\}\equiv\lbrack\phi\cdot\tilde{c}]_{1}.
\end{equation}
Whence, the natural representation of Diff$(\Sigma)$ on $\mathcal{V}%
_{\mathrm{LMI}}$ is given by%
\begin{equation}
\hat{U}(\phi)^{\prime}\Psi_{\lbrack\tilde{c}]_{i}}^{f^{i}}=\Psi_{\phi
^{-1}\cdot\lbrack\tilde{c}]_{i}}^{f^{i}\circ\phi}%
\end{equation}

Let us first evaluate the left hand side of (\ref{naiveattempt1}).
\begin{align}
\text{LHS}  &  =\left(  \hat{U}(\phi)^{\prime}\hat{H}^{1}[N]\hat{U}%
(\phi)^{-1\prime}\Psi_{\lbrack\tilde{c}]_{1}}^{f^{1}}\right)  \left(
|c_{A}\rangle\right) \label{naiveattempt2}\\
&  =\left(  \hat{H}^{1}[N]\hat{U}(\phi)^{-1\prime}\Psi_{\lbrack\tilde{c}]_{1}%
}^{f^{1}}\right)  \left(  |\phi\cdot c_{A}\rangle\right) \nonumber\\
&  =\left(  \hat{H}^{1}[N]^{\prime}\Psi_{\lbrack\phi\cdot\tilde{c}]}%
^{f^{1}\circ\phi^{-1}}\right)  |\phi\cdot c_{A}\rangle\nonumber\\
&  =\left[  \Psi_{\lbrack\phi\cdot\tilde{c}]}^{(f\circ\phi^{-1})_{\phi
(v_{0}),2}^{1,1}[N]}-\Psi_{\lbrack\phi\cdot\tilde{c}]}^{(f\circ\phi
^{-1})_{\phi(v_{0}),3}^{1,1}[N]}\right]  \left(  |\phi\cdot c_{A}%
\rangle\right) \nonumber
\end{align}
where, in the final line we have assumed (without loss of generality) that the
only vertex in $V(\phi\cdot\tilde{c})$ which lies in the support of $N$ is
$\phi(v_{0})$ with $v_{0}\in V(\tilde{c})$. The resulting vertex functions
$(f\circ\phi^{-1})_{\phi(v_{0}),2}^{1,1}[N]$ and $(f\circ\phi^{-1}%
)_{\phi(v_{0}),3}^{1,1}[N]$ are given by%
\begin{equation}
(f\circ\phi^{-1})_{\phi(v_{0}),2}^{1,1}[N]=\left\{
\begin{array}
[c]{ll}%
(f^{1}\circ\phi^{-1})\left(  \bar{V}(c^{\prime})\right)  , & \text{if}%
\ \phi\cdot v_{0}\ \notin\ \overline{V}(c^{\prime})\\
\lambda(\vec{n}_{\phi(v_{0})}^{\phi\cdot\tilde{c}})N(\phi(v_{0}))V_{2}%
^{a}(\phi(v_{0}))\partial_{a}^{\phi(v_{0})}(f^{1}\circ\phi^{-1})\left(
\phi(v_{0}),\bar{V}(c^{\prime})-\{\phi(v_{0})\}\right)  ,\quad &
\text{otherwise}%
\end{array}
\right.
\end{equation}
$(f\circ\phi^{-1})_{\phi(v_{0}),3}^{1,1}[N]$ is defined similarly with
$V_{2}^{a}$ replaced by $V_{3}^{a}$.\newline Whence, assuming $\phi\cdot
c_{A}\in\lbrack\phi\cdot\tilde{c}]_{1}$, we have
\begin{equation}
\text{LHS}=\left\{
\begin{array}
[c]{ll}%
0 & \text{if}\ \phi(v_{0})\notin\bar{V}(\phi\cdot c_{A})\\
N(\phi(v_{0}))\left[  V_{2}^{a}(\phi(v_{0}))-V_{3}^{a}(\phi(v_{0}))\right]
\partial_{a}^{\phi(v_{0})}(f^{1}\circ\phi^{-1})\left(  \phi(v_{0}),\bar
{V}(\phi\cdot\tilde{c})-\{\phi(v_{0})\}\right)  \quad & \text{if}\ \phi
(v_{0})\in\bar{V}(\phi\cdot c_{A})
\end{array}
\right.  \label{aug14-1-1}%
\end{equation}
and if $\phi\cdot c_{A}\notin\lbrack\phi\cdot\tilde{c}]_{1}$, we have
\begin{equation}
\text{LHS}=0 \label{aug14-2}%
\end{equation}
On the other hand, the right hand side of (\ref{naiveattempt1}) is given by%
\begin{equation}
\text{RHS}=\left(  \hat{H}^{1}[\phi^{\ast}N]^{\prime}\Psi_{\lbrack\tilde
{c}]_{1}}^{f^{1}}\right)  |c_{A}\rangle=\left[  \Psi_{\lbrack\tilde{c}]_{1}%
}^{f_{v_{0},2}^{1,1}[\phi^{\ast}N]}-\Psi_{\lbrack\tilde{c}]_{1}}^{f_{v_{0}%
,3}^{1,1}[\phi^{\ast}N]}\right]  |c_{A}\rangle\label{aug14-3}%
\end{equation}
where, following the assumption regarding the support of the lapse with
respect to the vertex set of $\phi\cdot\tilde{c}$, it is clear that the only
vertex in $V(\tilde{c})$ which lies inside the support of $\phi^{\ast}N$ is
$v_{0}$. As before, we can evaluate the resulting vertex functions, and find%
\begin{equation}
f_{v_{0},2}^{1,1}[\phi^{\ast}N](\bar{V}(c^{\prime}))=\left\{
\begin{array}
[c]{ll}%
f^{1}(V(c^{\prime})) & \text{if}\ v_{0}\notin\bar{V}(c^{\prime})\\
(\phi^{\ast}N)(v_{0})\lambda(\vec{n}_{v_{0}}^{\tilde{c}})V_{2}^{a}%
(v_{0})\partial_{a}^{v_{0}}f^{1}(v_{0},\bar{V}(c^{\prime})-\{v_{0}\})\quad &
\text{if}\ v_{0}\in\bar{V}(c^{\prime})
\end{array}
\right.  \label{aug14-4}%
\end{equation}
$f_{v_{0},3}^{1,1}[\phi^{\ast}N]$ is defined similarly with $V_{2}^{a}$
replaced by $V_{3}^{a}$.

Thus, if $c_{A}\in\lbrack\tilde{c}]_{1}$ we have%

\begin{equation}
\text{RHS}=\left\{
\begin{array}
[c]{ll}%
0 & \text{if}\ v_{0}\notin\bar{V}(c_{A})\\
(\phi^{\ast}N)(v_{0})\lambda(\vec{n}_{v_{0}}^{\tilde{c}})\left[  V_{2}%
^{a}(v_{0})-V_{3}^{a}(v_{0})\right]  \partial_{a}^{v_{0}}f^{1}(v_{0},\bar
{V}(c_{A})-\{v_{0}\})\quad & \text{if}\ v_{0}\in\bar{V}(c_{A})
\end{array}
\right.  \label{aug14-5-1}%
\end{equation}
and if $c_{A}\notin\lbrack\tilde{c}]_{1}$, we have%

\begin{equation}
\text{RHS}=0 \label{aug14-6}%
\end{equation}
We would now like to see if the LHS and RHS of (\ref{naiveattempt1}) as
detailed in (\ref{aug14-1-1}) to (\ref{aug14-5-1}) are equal for all
$\Psi_{\lbrack\tilde{c}]_{1}}^{f^{1}}$, $N,$ and $|c_{A}\rangle$.

\begin{enumerate}
\item[\textbf{Case 1}] $c_{A}\notin\lbrack\tilde{c}]_{1}\implies\phi\cdot
c_{A}\notin\lbrack\phi\cdot\tilde{c}]_{1}$. In this case from (\ref{aug14-2}),
(\ref{aug14-6}) we clearly see that LHS = RHS = 0.

\item[\textbf{Case 2}] $c_{A}\in\lbrack\tilde{c}]_{1}$ but $v_{0}\notin\bar
{V}(c_{A})\implies\phi\cdot c_{A}\in\lbrack\phi\cdot\tilde{c}]_{1}$ but
$\phi(v_{0})\notin\bar{V}(\phi\cdot c_{A})$. In this case from the first
equation in (\ref{aug14-1-1}) and in (\ref{aug14-5-1}) we see that LHS = RHS = 0.

\item[\textbf{Case 3}] $c_{A}\in\lbrack\tilde{c}]_{1}$ and $v_{0}\in\bar
{V}(c_{A})\implies\phi\cdot c_{A}\in\lbrack\phi\cdot\tilde{c}]_{1}$ and
$\phi(v_{0})\in\bar{V}(c_{A})$. In this case the LHS and RHS are given by the
first equations in (\ref{aug14-1-1}) and (\ref{aug14-5-1}) respectively. It is
clear that for a generic choice of the habitat state $\Psi_{\lbrack\tilde
{c}]_{1}}^{f^{1}}$, the LHS and RHS are not equal for two reasons:

\begin{enumerate}
\item[(i)] The LHS involves $N(\phi(v_{0}))$, whereas RHS involves
$(\phi^{\ast}N)(v_{0})=|\frac{d\phi}{dv_{0}}|^{\frac{1}{6}}N(\phi(v_{0}))$;
i.e., the two differ by a Jacobian factor.

\item[(ii)] The LHS involves $\lambda(\vec{n}_{\phi(v_{0})}^{\phi\cdot
\tilde{c}})\vec{V}_{i}(\phi(v_{0}),\phi\cdot\tilde{c})$ which is not equal to
$\lambda(\vec{n}_{v_{0}}^{\tilde{c}})\phi_{\ast}\vec{V}_{i}(v_{0},\tilde{c})$.
\end{enumerate}
\end{enumerate}

Thus the non-trivial density weight of the lapse and the diffeomorphism
non-covariance of the quantum shift are the two reasons why the naive attempt
to prove diffeomorphism covariance of $\hat{H}^{1}[N]$ fails.

\subsection{Checking Diffeomorphism-Covariance: $H^{2}[N]\Psi_{\lbrack
\tilde{c}]_{1}}^{f^{1}}$}

\label{NA-2}

In the previous section, we analyzed the behaviour of $\hat{H}^{1}[N]$ under
conjugation by the natural representation of Diff$\left(  \Sigma\right)  $ on
$\Psi_{\lbrack\tilde{c}]_{1}}^{f^{1}}$ and identified two problems which are
responsible for its spatial non-covariance. In this section, we continue along
the same route and analyze the diffeomorphism (non-)covariance of $\hat{H}%
^{2}[N]$ on $\Psi_{\lbrack\tilde{c}]_{1}}^{f^{1}}$. At the very least, we
expect the two culprits identified in the last section to spoil the covariance
properties again, but as we will see in this case there is an additional
difficulty. The action of $\hat{H}^{2}[N]$ on charge network states containing
EO vertices (which are by definition zero-volume) is different from its action
on non-EO zero-volume vertices. However, a quick look at the definition of EO
vertices reveals that the entire EO structure is diffeomorphism non-covariant:
A diffeomorphism can map an EO vertex into a WEO vertex. This transcends into
another issue in the continuum limit, ensuring diffeomorphism non-covariance
of $\hat{H}^{2}[N]$ on $\Psi_{\lbrack\tilde{c}]_{1}}^{f^{1}}$. We now turn to
a detailed analysis of this issue.

Given $\Psi_{\lbrack\tilde{c}]_{1}}^{f^{1}}\in\mathcal{V}_{\mathrm{LMI}}$ and
$|c_{A}\rangle\in\mathcal{H}_{\mathrm{kin}},$ we once again want to see if
\begin{equation}
\left(  \hat{U}(\phi)^{\prime}\hat{H}^{2}[N]\hat{U}(\phi^{-1})^{\prime}%
\Psi_{\lbrack\tilde{c}]_{1}}^{f^{1}}\right)  \left(  |c_{A}\rangle\right)
=\left(  \hat{H}^{(2)}[\phi^{\ast}N]\Psi_{\lbrack\tilde{c}]_{1}}^{f^{1}%
}\right)  \left(  |c_{A}\rangle\right)
\end{equation}
$\forall\phi\in\text{Diff}(\Sigma)$. We compute%
\begin{align}
\text{LHS}  &  =\left(  \hat{U}(\phi)^{\prime}\hat{H}^{2}[N]\hat{U}(\phi
^{-1})^{\prime}\Psi_{\lbrack\tilde{c}]_{1}}^{f^{1}}\right)  \left(
|c_{A}\rangle\right) \label{LHSofnaiveH2onS1}\\
&  =\left(  \hat{U}(\phi)^{\prime}\hat{H}^{2}[N]\Psi_{\phi\cdot\lbrack
\tilde{c}]_{1}}^{f^{1}\circ\phi^{-1}}\right)  \left(  |c_{A}\rangle\right)
\nonumber\\
&  =\left(  \hat{U}(\phi)^{\prime}\left[  \Psi_{\phi\cdot\lbrack\tilde{c}%
]_{1}}^{(f\circ\phi^{-1})_{\phi(v_{0}),1}^{1,2}[N]}-\Psi_{\phi\cdot
\lbrack\tilde{c}]_{1}}^{(f\circ\phi^{-1})_{\phi(v_{0}),3}^{1,2}[N]}\right]
\right)  \left(  |c_{A}\rangle\right) \nonumber\\
&  =\left[  \Psi_{\phi\cdot\lbrack\tilde{c}]_{1}}^{(f\circ\phi^{-1}%
)_{\phi(v_{0}),1}^{1,2}[N]}-\Psi_{\phi\cdot\lbrack\tilde{c}]_{1}}^{(f\circ
\phi^{-1})_{\phi(v_{0}),3}^{1,2}[N]}\right]  \left(  |\phi\cdot c_{A}%
\rangle\right)  ,\nonumber
\end{align}
where in the third line we have used (\ref{h2psi1})\ and assumed (without loss
of generality) that $\exists v_{0}\in V(\tilde{c})$ such that the only vertex
in $V(\phi\cdot\tilde{c})$ which falls inside the support of $N$ is
$\phi(v_{0})$. In the fourth line we have used (\ref{naturalrepdiff1}). The
vertex functions in the third and fourth lines of (\ref{LHSofnaiveH2onS1}) are
given in Section \ref{RECAP}, and
\begin{subequations}
\begin{align}
(f\circ\phi^{-1})_{\phi(v_{0}),3}^{1,2}[N]\left(  \bar{V}(c^{\prime})\right)
&  =\left(  f^{1}\circ\phi^{-1}\right)  (\bar{V}(c^{\prime}%
))\label{LHSofnaiveH2onS1-1}\\
(f\circ\phi^{-1})_{\phi(v_{0}),1}^{1,2}[N]\left(  \overline{V}(c^{\prime
})\right)   &  =\left(  f^{1}\circ\phi^{-1}\right)  (\bar{V}(c^{\prime}))
\end{align}
$\forall c^{\prime}$ such that $\bar{V}(c^{\prime})$ does not contain an EO
vertex $\phi(v_{0})_{\delta}^{\mathrm{E}}(1,2)$ of type$\left(  1,2\right)  $.

If on the other hand, $\overline{V}(c^{\prime})$ contains an EO vertex,
$(\phi\cdot v_{0})^{E}_{\delta}(1,2)$ (for some $\delta$) then,%

\end{subequations}
\begin{equation}
\label{LHSofnaiveH2onS1-2}%
\begin{array}
[c]{lll}%
(f\circ\phi^{-1})_{\phi\cdot v_{0}, 3}^{1,2}[N]\left(  \overline{V}(c^{\prime
})\right)  \ =(f\circ\phi^{-1})_{\phi\cdot v_{0}, 3}^{1,2}[N]\left(
(\phi\cdot v_{0})^{E}_{\delta}(1,2),\text{. . .}\right)  &  & \\
\vspace*{0.1in} \hspace*{0.5in}=\ \left[  \lambda(\vec{n}_{\phi\cdot v_{0}%
}^{\phi\cdot\tilde{c}})V^{a}_{1}(\phi\cdot v_{0},\phi\cdot\tilde{c}%
)(\nabla_{a}N)(\phi\cdot v_{0})\right]  f^{1}\circ\phi^{-1}\left(  \left(
\phi\cdot v_{0}\right)  ^{E}_{\delta}(1,3), \text{...}\right)  &  & \\
\vspace*{0.1in} (f\circ\phi^{-1})_{\phi\cdot v_{0}, 1}^{1,2}[N]\left(
\overline{V}(c^{\prime})\right)  \ =(f\circ\phi^{-1})_{\phi\cdot v_{0},
1}^{1,2}[N]\left(  (\phi\cdot v_{0})^{E}_{\delta}(1,2),\text{. . .}\right)  &
& \\
\vspace*{0.1in} \hspace*{0.5in}=\ \left[  \lambda(\vec{n}_{\phi\cdot v_{0}%
}^{\phi\cdot\tilde{c}})V^{a}_{3}(\phi\cdot v_{0},\phi\cdot\tilde{c}%
)(\nabla_{a}N)(\phi\cdot v_{0})\right]  f^{1}\circ\phi^{-1}\left(  \left(
\phi\cdot v_{0}\right)  ^{E}_{\delta}(1,1), \text{...}\right)  &  & \\
&  &
\end{array}
\end{equation}
Using the last line in (\ref{LHSofnaiveH2onS1}), it is easy to see that,
\begin{equation}
\label{LHSofnaiveH2onS1-3}%
\begin{array}
[c]{lll}%
\text{LHS}\ =\ 0\ \text{if}\ \phi\cdot c_{A}\ \notin\ \phi\cdot[c]_{1} &  & \\
\vspace*{0.1in} \hspace*{0.5in}=\left[  (f\circ\phi^{-1})_{\phi\cdot v_{0},
1}^{1,2}[N]\left(  \overline{V}(c_{A})\right)  \ -\ (f\circ\phi^{-1}%
)_{\phi\cdot v_{0}, 3}^{1,2}[N]\left(  \overline{V}(c_{A})\right)  \right]
\ \text{otherwise} &  &
\end{array}
\end{equation}
Whence upon using (\ref{LHSofnaiveH2onS1-2}) in (\ref{LHSofnaiveH2onS1-3}) we
see that if $c_{A}$ does not contain an EO vertex of type-$1,2$ associated to
$v_{0}$ then
\begin{equation}
\label{LHSofnaiveH2onS1-4}\text{LHS}\ =\ 0
\end{equation}

We now turn our attention to RHS $\left(  \hat{H}^{(2)}[\phi^{*}N]\Psi^{f^{1}%
}_{[\tilde{c}]_{1}}\right)  \vert c_{A}\rangle$.\newline%
\begin{equation}
\label{RHSofnaiveH2onS1}%
\begin{array}
[c]{lll}%
\text{RHS}\ =\ \left[  \Psi^{f_{v_{0},1}^{1,2}[\phi^{*}N]}_{[\tilde{c}]_{1}%
}\ -\ \Psi^{f_{v_{0},3}^{1,2}[\phi^{*}N]}_{[\tilde{c}]_{1}}\right]  \vert
c_{A}\rangle &  & \\
&  &
\end{array}
\end{equation}
where the vertex functions are once again given by,%

\begin{equation}
\label{sep3-2}%
\begin{array}
[c]{lll}%
f_{v_{0},3}^{1,2}[\phi^{*}N]\left(  \overline{V}(c^{\prime})\right)  \ = &  &
\\
\vspace*{0.1in} \hspace*{0.6in} f_{1}\left(  \overline{V}(c^{\prime})\right)
\ \text{if there is no EO vertex of type-1,2 w.r.t}\ v_{0}\ \text{in}
\overline{V}(c^{\prime}) &  & \\
\vspace*{0.1in} \hspace*{0.6in} \left[  \lambda(\vec{n}_{v_{0}}^{\tilde{c}%
}\ V^{a}_{1}(\tilde{c}, v_{0})(\nabla_{a}\left(  \phi^{*}N\right)
)(v_{0})\right]  f^{1}\left(  \left(  v_{0}\right)  ^{E}_{\delta}(1,3),
\text{. . .}\right)  \ \text{if there is an E.O vertex} (v_{0})^{E}_{\delta}
\text{w.r.t} v_{0} &  & \\
\vspace*{0.1in} f_{v_{0},1}^{1,2}[\phi^{*}N]\left(  \overline{V}(c^{\prime
})\right)  \ = &  & \\
\vspace*{0.1in} \hspace*{0.6in} f_{1}\left(  \overline{V}(c^{\prime})\right)
\ \text{if there is no EO vertex of type-1,2 w.r.t}\ v_{0}\ \text{in}
\overline{V}(c^{\prime}) &  & \\
\vspace*{0.1in} \hspace*{0.6in} \left[  \lambda(\vec{n}_{v_{0}}^{\tilde{c}%
}\ V^{a}_{3}(\tilde{c}, v_{0})(\nabla_{a}\left(  \phi^{*}N\right)
)(v_{0})\right]  f^{1}\left(  \left(  v_{0}\right)  ^{E}_{\delta}(1,1),
\text{. . .}\right)  \ \text{if there is an E.O vertex} (v_{0})^{E}_{\delta}
\text{w.r.t} v_{0} &  & \\
&  &
\end{array}
\end{equation}

Thus it is straight-forward to see that, RHS as defined in
(\ref{RHSofnaiveH2onS1}) is given by,
\begin{equation}
\label{sep3-1}%
\begin{array}
[c]{lll}%
\text{RHS}\ =\ 0\ \text{if}\ c_{A}\ \notin\ [\tilde{c}]_{1} &  & \\
\vspace*{0.1in} \hspace*{0.6in}=\ f_{v_{0},1}^{1,2}[\phi^{*}N]\left(
\overline{V}(c_{A})\right)  \ -\ f_{v_{0},3}^{1,2}[\phi^{*}N]\left(
\overline{V}(c_{A})\right)  \ \text{otherwise} &  &
\end{array}
\end{equation}

which using (\ref{sep3-2}) further implies that%

\begin{equation}
\label{sep4-1}%
\begin{array}
[c]{lll}%
\text{RHS}\ =\ 0\ \text{if there exists no E.O. vertex of type-I,j
in}\ V(c_{A}) \text{associated to}\ v_{0}. &  & \\
\vspace*{0.1in} \hspace*{0.6in}\ =\ -\lambda(\vec{n}_{v_{0}}^{\tilde{c}%
})\ \left[  V^{a}_{3}(\tilde{c}, v_{0}) f^{1}\left(  (v_{0})^{E}_{\delta
}(1,1), \text{. . .}\right)  -V^{a}_{1}(\tilde{c}, v_{0}) f^{1}\left(
(v_{0})^{E}_{\delta}(1,3), \text{. . .}\right)  \right]  \ (\nabla_{a}\left(
\phi^{*}N\right)  )(v_{0}) &  & \\
\hspace*{5.0in} \text{otherwise} &  &
\end{array}
\end{equation}

Comparing (\ref{sep4-1}) with (\ref{LHSofnaiveH2onS1-3}) we can easily verify
that, if $c_{A}\ \notin\ [\tilde{c}]_{1}$ (which is equivalent to $\phi\cdot
c_{A}\ \notin\ [\phi\cdot\tilde{c}]_{1}$) then%

\begin{equation}%
\begin{array}
[c]{lll}%
\text{LHS}\ =\ \text{RHS}\ =\ 0 &  &
\end{array}
\end{equation}

However if $c_{A}\ \in\ [\tilde{c}]_{1}$ then LHS and RHS are only equal for
all diffeomorphisms, if there is no WEO vertex of type-$(1,2)$ associated to
$v_{0}$ in $c_{A}$. Otherwise there could exist a diffeomorphism $\phi$ such
that it would map a WEO vertex associated to $v_{0}$ to an EO vertex (of the
same type) associated to $\phi\cdot v_{0}$ in which case LHS would be zero
(from (\ref{LHSofnaiveH2onS1-4}) but RHS would be given by the second line in
(\ref{sep4-1}).\newline

We thus conclude that given a $\Psi^{f_{i}}_{[c]_{i}}$ in the LMI-habitat, the
action of $\hat{H}[N]\ =\ \sum_{j=1}^{3} H^{(j)}[N]^{\prime}$ is not covariant
under action of spatial diffeomorphisms due to three reasons.\newline

\noindent{(a)} The quantum shift is not a covariant object in any sense : If
two charge-networks $c_{1}$ and $c_{2}$ are diffeomorphic to each other (which
means there are infinitely many semi-analytic diffeomorphisms which map
$c_{1}$ to $c_{2}$; there need not exist any diffeomorphism whose push-forward
maps $\vec{V}_{j}(v, c_{1})$ to $\vec{V}_{j}(\phi\cdot v, c_{2})$%
.\newline\noindent{(b)} The non-trivial density weight of lapse causes extra
Jacobian factors to arise when comparing $\phi^{*}\ N$ with $N\circ\phi
$.\newline\noindent{( c )} The EO structure is a diffeomorphism non-covariant
concept unlike the WEO structure.

\subsection{Our Strategy}

In this section we briefly outline our approach and explain the key ideas that
are developed in subsequent sections. As some of the analysis done in later
sections is slightly involved, we hope that a reading of this section will
give the reader an understanding of the concepts.\newline Our aim is to show
that despite the apparent background dependence of quantum Hamiltonian
constraint , we obtain an anomaly free representation of the Dirac algebra on
$\mathcal{V}_{LMI}$ by defining a new representation of the diffeomorphism
group on $\mathcal{H}_{kin}$ (and whence by dual action on $\mathcal{V}_{LMI}%
$). The basic ideas behind our construction are summarized below.\newline As
shown in \cite{hat}, and recalled briefly in the Section \ref{RECAP}, the
continuum Hamiltonian constraint on the LMI habitat, is a sum of three
operators given by,
\begin{equation}%
\begin{array}
[c]{lll}%
\hat{H}[N]\Psi_{\lbrack\tilde{c}]_{i}}^{f^{i}}\ =\ \left(  \hat{H}%
^{1}[N]\ +\ \hat{H}^{2}[N]\ +\ \hat{H}^{3}[N]\right)  \Psi_{\lbrack\tilde
{c}]_{i}}^{f^{i}}\ \forall\ i\in\{1,2,3\} &  &
\end{array}
\end{equation}
Where $\Psi_{\lbrack\tilde{c}]_{i}}^{f^{i}}$ is an arbitrary element in
$\mathcal{V}_{LMI}$. We will restrict our analysis to $i=1$ case (as the
analysis for states in the $i=2,3$ sectors is exactly analogous) and prove
diffeomorphism covariance of $H[N]$ by showing
\begin{equation}%
\begin{array}
[c]{lll}%
\hat{U}(\phi)\hat{H}^{1}[N]\hat{U}(\phi^{-1})\Psi_{\lbrack\tilde{c}]_{1}%
}^{f^{1}}\ =\ \hat{H}^{1}[\phi^{\ast}N]\Psi_{\lbrack\tilde{c}]_{1}}^{f^{1}} &
& \\
\vspace*{0.1in}\hat{U}(\phi)\hat{H}^{j}[N]\hat{U}(\phi^{-1})\Psi
_{\lbrack\tilde{c}]_{1}}^{f^{1}}\ =\ \hat{H}^{j}[\phi^{\ast}N]\Psi
_{\lbrack\tilde{c}]_{1}}^{f^{1}}\ j\in\{2,3\} &  &
\end{array}
\label{diffcovaim}%
\end{equation}
From eq. (\ref{diffcovaim}) it follows that $\hat{H}[N]$ is
diffeomorphism-covariant on any $\Psi_{\lbrack\tilde{c}]_{1}}^{f^{1}}%
\in\mathcal{V}_{\mathrm{LMI}}$. Diffeomorphism-covariance of $\hat{H}[N]$ on
an arbitrary state in $\mathcal{V}_{LMI}$ is a trivial extension of the above
claim.\newline

We now describe the main ideas behind the new representation of Diff$(\Sigma)$
defined in Section \ref{newrep}. As recalled in Section \ref{RECAP}, given any
charge network $c$, there is a unique \textquotedblleft
undeformed\textquotedblright\ $\tilde{c}$ associated to it such that the
action of the finite-triangulation Hamiltonian constraint on $c$ involves a
set of vectors $\vec{V}(v,\tilde{c})$ associated to each vertex $v\in V(\tilde{c}),$
and a characterization of which of the vertices $v^{E}(i,j)$ in $c$ are EO with respect to
vertices in $\tilde{c}$. Whence it is clear that the data set we are dealing
with, as far as the definition of the Hamiltonian constraint action on $c$
goes, is $\{\vec{V}(v,\tilde{c}),v^{\mathrm{E}}(i,j)|v\in V(\tilde{c})\cap V( c ),\ v^{\mathrm{E}}(i,j)\ \in V( c )\}$. 
Denote the collection of all such data sets associated to any diffeomorphism invariant orbit of charge networks by $\mathcal{C}([c]_{diff})\ =\cup_{c^{\prime}\ \in [c]_{diff}}%
\{\vec{V}(v,\tilde{c}^{\prime}),v^{\mathrm{E}}(i,j)|v\in\ V(\tilde{c}^{\prime}),\ v^{\mathrm{E}}(i,j)\ \in V( c^{\prime} )\}$. Intuitively
we would like to choose a representation of Diff$(\Sigma)$ on $\mathcal{H}%
_{\mathrm{kin}}$ which preserves $\mathcal{C}([c]_{diff})$; that is, any diffeomorphism
should act in such a way that it maps one element of $\mathcal{C}([c]_{diff})$ to some
other element of $\mathcal{C}([c]_{diff})$). We achieve this objective as follows.

\noindent{(i)} In each diffeomorphism invariant orbit $[\tilde{c}]_{\mathrm{diff}}$ of undeformed
charge networks, we fix once and for all, an \textquotedblleft
initial\textquotedblright\ charge network $\tilde{c}^{0}\in\lbrack\tilde
{c}]_{\mathrm{diff}}\equiv\lbrack\tilde{c}^{0}]_{\mathrm{diff}},$ and a set of
diffeomorphisms $\{\phi_{\tilde{c}^{0},\tilde{c}^{\prime}}\}_{\tilde
{c}^{\prime}\in\lbrack\tilde{c}^{0}]_{\mathrm{diff}}}$ which map $\tilde
{c}^{0}$ to any $\tilde{c}^{\prime}\in\lbrack\tilde{c}^{0}]_{\mathrm{diff}}$.\\
\noindent{(ii)} We also associate to each such $\tilde{c}^{0}$ an atlas $\mathcal{U}(\tilde
{c}^{0})$ on $\Sigma$ such that each vertex $v$ of $\tilde{c}^{0}$ lies in
precisely one open set of $\mathcal{U}(\tilde{c}^{0})$ and to each $\tilde
{c}^{\prime}\in\lbrack\tilde{c}^{0}]_{\mathrm{diff}}$ we associate an atlas
obtained by pushforward\footnote{By pushforward of the coordinate chart we
merely mean the coordinates of the diffeomorphic image of a point in the
pushed-forward coordinate chart are the same as the coordinates of original
point in the initial coordinate chart.} of $\mathcal{U}(\tilde{c}^{0})$ by
$\phi_{\tilde{c}^{0},\tilde{c}^{\prime}}$.\\ 
\noindent{(iii)} We compute the quantum shift vectors $\{\vec{V}(v,\tilde{c}^{0})|v\in
V(\tilde{c}^{0})\}$ on the vertices of reference charge-network $\tilde{c}%
^{0}$ once and for all and \emph{define} quantum shift vectors for any
$\tilde{c}^{\prime}\in\lbrack\tilde{c}^{0}]_{\mathrm{diff}}$ as%
\begin{equation}
\vec{V}(v^{\prime},\tilde{c}^{\prime}):=(\phi_{\tilde{c}^{0},\tilde{c}%
^{\prime}})_{\ast}\vec{V}(\phi_{\tilde{c}^{0},\tilde{c}^{\prime}}%
^{-1}(v^{\prime}),\tilde{c}^{0})
\end{equation}
Thus, given a $[\tilde{c}]_{\mathrm{diff}}$ with a reference charge network
$\tilde{c}^{0}$, the set $\mathcal{C}([c]_{diff})$ (such that the unique WEO-vertex free charge network associated to $c$ is $\tilde{c}$,) is invariant under the action of
diffeomorphisms $\phi_{\tilde{c}^{0},\tilde{c}}$. This motivates our new
representation which essentially amounts to working with $\cup_{\tilde{c}^{0}%
}\cup_{\tilde{c}^{\prime}\in\lbrack\tilde{c}^{0}]_{\mathrm{diff}}}\phi
_{\tilde{c}^{0},\tilde{c}^{\prime}}$ instead of $\mathrm{Diff}(\Sigma
)$.\footnote{At this point we are only trying to motivate our construction of
new representation. The details are given in Section \ref{newrep}. For
example, at this point we have not even shown that the set that we are working
with forms a group.} We now revisit the transformation properties of $\hat
{H}[N]$ on $\mathcal{V}_{\mathrm{LMI}}$ under this representation of
Diff$\left(  \Sigma\right)  $ and show that it transforms covariantly.

\subsection{Preferred diffeomorphisms : $\phi$-maps}\label{phimaps}
In this section we explain how we assign to each diffeomorphism-invariant orbit $[\tilde{c}]_{diff}$ of WEO-vertex free charge networks a set of diffeomorphisms which will be a crucial ingredient in defining a new representation of $Diff(\Sigma)$ in the quantum theory.\\
We start with a trivial observation. $[\tilde{c}]_{diff}$ is a category (in fact a groupoid) with 
$\tilde{c}^{\prime}\ \in\ [\tilde{c}]_{diff}$ being the objects and all the diffeomorphisms which map say $\tilde{c}^{\prime}$ to $c^{\prime\prime}$ constitute $Hom(\tilde{c}^{\prime},\tilde{c}^{\prime\prime})$. Our idea is to work with a  subcategory (in fact a subgroupoid) in $[\tilde{c}]_{diff}$ in defining a representation of $Diff(\Sigma)$ on ${\cal H}_{kin}$. Pick a reference charge-network $\tilde{c}_{0}$ and for all $\tilde{c}^{\prime}\ \in\ [\tilde{c}]_{diff}$ fix once and for all a diffeomorphism $\phi_{\tilde{c}_{0},\tilde{c}^{\prime}}$ which map $\tilde{c}^{0}$ to $\tilde{c}^{\prime}$. 
(We choose $\phi_{\tilde{c}_{0},\tilde{c}^{0}}\ =\ \textrm{Id}$). Now given any $\tilde{c}^{\prime}, \tilde{c}^{\prime\prime}\ \in [\tilde{c}]_{diff}$ 
we define a diffeomorphism which maps $\tilde{c}^{\prime}$ to $\tilde{c}^{\prime\prime}$ as
\begin{equation}
\begin{array}{lll}
\phi_{\tilde{c}^{\prime}, \tilde{c}^{\prime\prime}}\ :=\ \phi_{\tilde{c}^{0},\tilde{c}^{\prime\prime}}\circ\phi_{\tilde{c}^{0},\tilde{c}^{\prime}}^{-1}
\end{array}
\end{equation}
It is easy to verify that
\begin{equation}
\begin{array}{lll}
\phi_{\tilde{c}_{1},\tilde{c}_{2}}\ \circ\ \phi_{\tilde{c}_{2},\tilde{c}_{3}}\ =\ \phi_{\tilde{c}_{1},\tilde{c}_{3}}\\
\vspace*{0.1in}
\phi_{\tilde{c}_{1},\tilde{c}_{2}}\ =\ \phi_{\tilde{c}_{2},\tilde{c}_{1}}^{-1}
\end{array}
\end{equation}
$\forall\ \tilde{c}_{1},\tilde{c}_{2}\ \in\ [\tilde{c}]_{diff}$.\\
The categorical notions are not essential in understanding the representation of $Diff(\Sigma)$, however it is a useful concept to understand the type of structure we are dealing with when we fix a diffeomorphism once and for all between any two charge-networks.\\
We will sometimes refer to these select set of diffeomorphisms as $\phi$-maps.

\subsection{A Diffeomorphism-Covariant Regularization Scheme}

\label{diffcovquantumshift}


Classically, $V_{i}^{a}=q^{-1/4}E_{i}^{a}$ is a $C^{\infty}$ densitized vector
field. In \cite{hat} the quantization of $V_{i}^{a}(v)$ at a given point
$v\in\Sigma$ involved a choice of regulating structures such that, at finite
regularization parameterized by $\epsilon$, a densely defined operator
$\hat{V}_{i}^{a}(v)|_{\epsilon}$ on $\mathcal{H}_{\mathrm{kin}}$ was obtained.
Although this operator is explicitly independent of $\epsilon$ due to the
density weights of various quantities, it is implicitly dependent on the
chosen regulating structures, which can be most easily seen through its
spectrum. In particular, this dependence implies that generically, given two
charge-networks $c_{1}$ and $c_{2}$ that are diffeomorphic to each other,
\begin{equation}
\phi_{\ast}\left(  \langle c|\hat{\vec{V}}_{i}(v)|_{\epsilon}|c_{1}%
\rangle\right)  \neq\langle\phi c|\hat{\vec{V}}_{i}(\phi\cdot v)|_{\epsilon}%
|c_{2}\rangle. \label{noncov}%
\end{equation}
$\forall\ \phi$ which map $c_{1}$ to $c_{2}$. This result implies the following.

Consider $[c_{0}]_{\mathrm{diff}}$ which is a diffeomorphism-invariant set of
charge networks that contains $c_{0}$. The defintion of the quantum shift
vectors associated to a given $c$ is essentially an assignment of vectors
$\{\vec{V}_{i}(v, c)|v\in V(c),\ i\in\{1,2,3\}\}$. Eq. (\ref{noncov}) implies
that, given $c_{1},c_{2}\in\lbrack c_{0}]_{\mathrm{diff}},$, there \emph{is no} meaningful sense in which we can talk about the quantum shift vectors associated to $c_{1}$ being diffeomorphically related to quantum shift vectors associated to $c_{2}$. We term this
property, diffeomorphism non-covariance of quantum shift.
As we saw in Section \ref{NA}, the diffeomorphism non-covariance of
the quantum shift in turn implies that the Hamiltonian constraint operator as
we have defined it will not be diffeomorphism-covariant; i.e., Equation
(\ref{diracalg2}) will not be satisfied. We cure this problem by first taking
a cue from the construction of rigging map for finite diffeomorphisms \cite{almmt}, then
defining an alternative (as opposed to the representation currently used in LQG) representation of Diff$\left(  \Sigma\right)  $ on
$\mathcal{H}_{\mathrm{kin}}$. We show that this leads to a
solution to the diffeomorphism non-covariance problem of the quantum shift,\footnote{This means that we can in a precise sense talk about a map between quantum shifts associated to two charge-networks which are diffeomorphic to each other} and finally to a 
diffeomorphism-covariant Hamiltonian constraint operator.\newline

First let us\ briefly recall the result of the construction of the quantum
shift in \cite{hat}. At each point $p\in\Sigma,$ we fix once and for all a
coordinate system $\{x_{p}\}$ with origin at $p.$ Let $\tilde{c}$ be a WEO vertex-free charge network
with a vertex $v\in\Sigma.$ The (co-ordinate dependent) regularization procedure, detailed in
\cite{hat} gives%
\begin{equation}
\boxed{ \hat{V}_{j}^{a}(v)|_{\epsilon}|\tilde{c}\rangle=\hat{E}_{j}^{a}|_{\epsilon}(v)\hat{q}_{\epsilon}^{-1/4}(v)|\tilde{c}\rangle= \lambda(\vec{n}_{\tilde{c}}^{v})\frac {1}{\pi}\sum_{e_{I}\cap v}\hat{e}_{I}^{a}n_{I}^{j}|\tilde{c}\rangle\equiv \lambda(\vec{n}_{\tilde{c}}^{v})\langle E_{j}^{a}(v)\rangle_{\tilde{c}}|\tilde{c}\rangle =:V_{j}^{a}(v,\tilde{c})|\tilde{c}\rangle }
\end{equation}
where  we have employed the abstract index notation and the $\hat{e}_{I}^{a}$ are unit tangent vectors (with respect to the
coordinate system at $v$ with metric $\delta_{ab}$) to the edges of $\tilde{c}$
emanating from $v.$
As it stands, $V_{j}^{a}(v,\tilde{c})$ is computed separately for each member $\tilde{c}$ of
the diffeomorphism equivalence class $[\tilde{c}]_{\mathrm{diff}}$ of $\tilde{c}.$

To solve the non-covariance problem stated above, we will modify this
construction, and compute $V_{j}^{a}(v,\tilde{c}_{0})$ only in some reference
charge network $\tilde{c}_{0}\in\lbrack \tilde{c}]_{\mathrm{diff}}\equiv\lbrack \tilde{c}_{0}%
]_{\mathrm{diff}}.$ The result will be transported to vertices of other
charge-nets in the equivalence class by the set of relevant $\phi$-maps which were defined in \ref{phimaps}.\\
We define the quantum shift vectors $V_{i}^{a}(v,\tilde{c})\ \forall\ v :=\ \phi_{\tilde{c}_{0},\tilde{c}}\cdot v_{0}|v\ \in\ V(\tilde{c})$, $\tilde{c}\in\lbrack
\tilde{c}_{0}]_{\mathrm{diff}}$ via pushforward with respect to the $\varphi_{\tilde{c}_{0},\tilde{c}}$:%
\begin{equation}
\begin{array}{lll}
\vec{V}_{j}(v,\tilde{c})\ :=\ (\phi_{\tilde{c}_{0},\tilde{c}})_{*}\vec{V}_{j}(v_{0},\tilde{c}_{0})
\end{array}
\end{equation}

\bigskip

\section{Representation of $Diff(\Sigma)$ on the LMI Habitat}

\label{newrep}
As we saw in Section \ref{NA-1}, there is a natural representation of
Diffeomorphism group on $\mathcal{V}_{\mathrm{LMI}}$. It is given by,
\begin{equation}
\hat{U}(\phi)^{\prime}\Psi_{\lbrack c]_{i}}^{f^{i}}=\psi_{\lbrack\phi\cdot
c]_{i}}^{f^{i}\circ\phi^{-1}}%
\end{equation}
However as we saw in Section \ref{NA-2}, this representation is not the one
that will lead us to a non-anomalous Dirac algebra on $\mathcal{V}%
_{\mathrm{LMI}}$, as $\hat{U}(\phi)$ generically maps an EO vertex to a WEO
vertex. Keeping this in mind, we define a new representation of Diff$(\Sigma)$
on $\mathcal{H}_{\mathrm{kin}}$ which in turn leads to a novel representation
of the diffeomorphism group on $\mathcal{V}_{\mathrm{LMI}}$. We will see that
this representation has some desirable properties.

\begin{enumerate}
\item[(1)] Given $[\tilde{c}_{0}]_{\mathrm{diff}}$, and the collection of vectors
$\cup_{\tilde{c}^{\prime}\in\lbrack \tilde{c}_{0}]_{\mathrm{diff}}}\{\vec{V}_{i}(v,\tilde{c}^{\prime
})|v\in V(\tilde{c}^{\prime})\}$, the new representation preserves this set. More
precisely, $(\phi_{\tilde{c}_{0},\tilde{c}^{\prime}})_{\ast v}\left(  \vec{V}_{i}%
(v,\tilde{c}_{0})\right)  =\vec{V}_{i}(v^{\prime},\tilde{c}^{\prime})\ $for all$\ v\in
V(\tilde{c}_{0})\ \text{such that }v^{\prime}=\phi_{\tilde{c}_{0},\tilde{c}^{\prime}}(v)\in
V(\tilde{c}^{\prime})$.

\item[(2)] It preserves the EO structure associated to charge-nets. (As we
will see below, this will be achieved by making co-ordinate charts around a
given vertex \textquotedblleft state dependent").
\end{enumerate}

\subsection{An Alternative Representation of $\mathrm{Diff}(\Sigma)$}

\label{newrep-1}

\subsubsection{Preliminaries}

\emph{Definitions:}\ Let $\tilde{c}$ be a WEO vertex-free (signified by the
tilde) charge network with vertex set $V(\tilde{c})=:\{v_{1},\dots$ $,v_{I},$
$\dots$ $v_{|V(\tilde{c})|}\},$ and let $\delta<\delta_{0}(\tilde{c})$ be an
admissible small parameter with respect to each of the coordinate systems
based at the points of $V(\tilde{c}),$ as detailed in \cite{hat} (roughly, the
bound $\delta_{0}(\tilde{c})$ guarantees that the finite-triangulation
Hamiltonian-type deformations at `fineness' $\delta$ that are performed on
$\tilde{c}$ are `local enough' so that one can actually classify these
so-called EO vertices which are formed by the action of Hamiltonian
constraint). We define the $i^{\mathrm{th}}$ $\delta$\emph{-cilium} at the
vertex $v_{I},$ denoted $\sigma_{i}^{I}(\delta,\tilde{c}),$ as a straight-line
arc of coordinate length $\delta|\langle\hat{E}_{i}^{a}\rangle_{\tilde{c}}|,$
directed along the quantum shift $V_{i}^{a}(v_{I},\tilde{c})$, with one end at
$v_{I},$ which goes into the definition of the curvature loop appearing in the
Hamiltonian action.

\begin{figure}
\begin{center}
\includegraphics[scale=0.8]{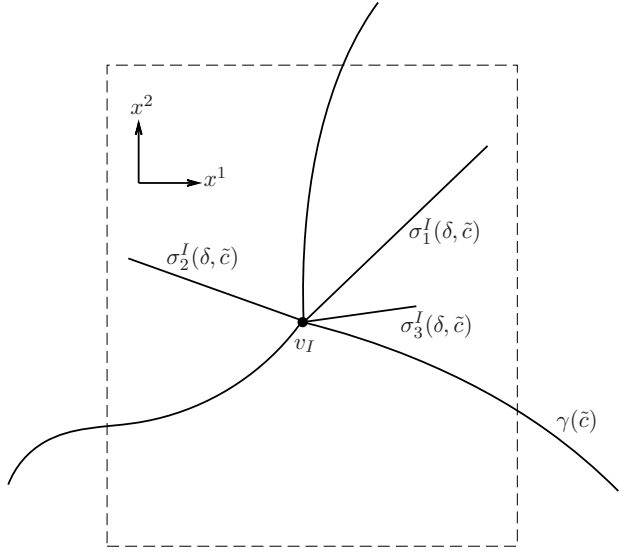}
\caption{The $\delta$-ciliated graph $\gamma_{\sigma(\delta,\tilde{c})}%
(\tilde{c})$ determined by $\tilde{c}%
$ in the neighborhood of the vertex $v_I$ with respect to the coordinate system $\{x_{v_I}%
\}=\{x^1,x^2\}$.}
\label{ciliatedfig}
\end{center}
\end{figure}

The $\delta$\emph{-ciliated graph} determined by $\tilde{c},$ denoted
$\gamma_{\sigma(\delta,\tilde{c})}(\tilde{c}),$ is given by the union of the
graph $\gamma(\tilde{c})$ underlying $\tilde{c}$ and the set of $\delta
$-cilia (see Figure (\ref{ciliatedfig})):
\begin{equation}
\gamma_{\sigma(\delta,\tilde{c})}(\tilde{c}):=\gamma(\tilde{c})\cup
\bigcup_{i=1}^{3}\bigcup_{I=1}^{|V(\tilde{c})|}\sigma_{i}^{I}(\delta,\tilde
{c}).
\end{equation}
Now consider the diffeomorphism equivalence class $[\tilde{c}]_{\mathrm{diff}%
}$. We choose once and for all a preferred element of $\tilde{c}_{0}\in
\lbrack\tilde{c}]_{\mathrm{diff}}\equiv\lbrack\tilde{c}_{0}]_{\mathrm{diff}}$
to represent the equivalence class, and in the neighborhood of each $v\in
V(\tilde{c}_{0})$ a fixed coordinate chart $\{x_{v}\}^{\tilde{c}_{0}}$. For
all $\tilde{c}\in\lbrack\tilde{c}_{0}]_{\mathrm{diff}},$ we choose once and
for all a preferred collection of coordinate charts $\{x_{v^{\prime}%
}\}^{\tilde{c}}$ in the neighborhood of each of its vertices $v^{\prime}%
=\phi_{\tilde{c}_{0},\tilde{c}}(v)$ which is obtained by a push-forward of
$\{x_{v}\}^{\tilde{c}_{0}}$ using $\phi_{\tilde{c}_{0},\tilde{c}}$.%
\begin{equation}%
\begin{array}
[c]{lll}%
\{x_{v^{\prime}}\}^{\tilde{c}}\ =\ (\phi_{\tilde{c}_{0},\tilde{c}})_{\ast
}\{x_{v}\}^{\tilde{c}_{0}} &  & \\
\vspace*{0.1in}\forall v^{\prime}\in V(\tilde{c}),\ v\in V(\tilde{c}%
_{0})\ \text{such that}\ v^{\prime}=\phi_{\tilde{c}_{0},\tilde{c}}(v) &  &
\end{array}
\end{equation}
This means that given a $v_{0}\in V(\tilde{c}_{0})$, $\sigma_{i}(v_{0}%
,\delta,\tilde{c}_{0})$ which is a linear curve (in parameter $t\in
\lbrack0,\delta]$) with respect to the coordinate chart $\{x_{v}\}^{\tilde
{c}_{0}}$, beginning at $v$ with its tangent at $v$ being $\vec{V}%
_{i}(v,\tilde{c}_{0})$ gets mapped to a linear curve $\sigma_{i}(v^{\prime
},\delta,\tilde{c})$ (in parameter $t\in\lbrack0,\delta]$) with respect to the
co-ordinate chart $\{x_{v^{\prime}}\}^{\tilde{c}}$, beginning at $v$ with its
tangent at $v$ being $\vec{V}_{i}(v^{\prime},\tilde{c}):=(\phi_{\tilde{c}%
_{0},\tilde{c}})_{\ast}[\vec{V}_{i}(v,\tilde{c}_{0})]$. Whence, we get%

\begin{align}
\phi_{\tilde{c}_{0},\tilde{c}}\left(  \gamma_{\sigma(\delta,\tilde{c}_{0}%
)}(\tilde{c}_{0})\right)   &  =\phi_{\tilde{c}_{0},\tilde{c}}\left(
\gamma(\tilde{c}_{0})\cup%
{\textstyle\bigcup\nolimits_{i=1}^{3}}
{\textstyle\bigcup\nolimits_{I=1}^{|V(\tilde{c}_{0})|}}
\sigma_{i}^{I}(\delta,\tilde{c}_{0})\right) \label{preserveEO-1}\\
&  =\gamma(\tilde{c})\cup%
{\textstyle\bigcup\nolimits_{i=1}^{3}}
\phi_{\tilde{c}_{0},\tilde{c}}\left(
{\textstyle\bigcup\nolimits_{I=1}^{|V(\tilde{c}_{0})|}}
\sigma_{i}^{I}(\delta,\tilde{c}_{0})\right) \nonumber\\
&  =\gamma(\tilde{c})\cup%
{\textstyle\bigcup\nolimits_{i=1}^{3}}
\left(
{\textstyle\bigcup\nolimits_{I=1}^{|V(\tilde{c}_{0})|}}
\sigma_{i}^{I}(\delta,\tilde{c})\right) \nonumber\\
&  =\gamma_{\sigma(\delta,\tilde{c})}(\tilde{c})\nonumber
\end{align}
Whence by adapting the coordinate charts around vertices to charge networks, we
preserve the EO nature of a vertex.\footnote{We are indebted to Madhavan Varadarajan who explained to us the use of state-dependent co-ordinate charts in regularization of quantum constraints.} As we will see, this new ingredient will
turn out to be crucial in obtaining an anomaly-free constraint algebra.

\subsubsection{A New Representation}

We are now ready to define a new representation of $\text{Diff}(\Sigma)$ on
$\mathcal{V}_{\mathrm{LMI}}$ via a new representation on $\mathcal{H}%
_{\mathrm{kin}}$. For charge networks $\tilde{c}$ with no WEO vertices, we
have that%
\begin{equation}
\hat{U}(\phi)|\tilde{c}\rangle:=|\phi_{\tilde{c},\phi\tilde{c}}\cdot\tilde
{c}\rangle.
\end{equation}
If a charge network $c_{1}$ has any WEO vertices, then, as shown in the
appendix, there is a unique WEO vertex-free charge network $\tilde{c}(c_{1})$
associated to it
and $\tilde{c}(c_{1})$ can be recovered from $c_{1}$ by performing a certain
surgery. Using this fact we then define%
\begin{equation}
\hat{U}(\phi)|c\rangle:=|\phi_{\tilde{c}(c),\phi\cdot\tilde{c}(c)}\cdot
c\rangle.
\end{equation}
Clearly this defines a representation:%
\begin{equation}
\hat{U}(\phi^{\prime})\hat{U}(\phi)|c\rangle=|\phi_{\widetilde{\phi_{\tilde
{c},\phi\cdot\tilde{c}}\cdot c},\phi^{\prime}\cdot\widetilde{\phi_{\tilde
{c},\phi\cdot\tilde{c}}c}}\phi_{\tilde{c},\phi\cdot\tilde{c}}\cdot
c\rangle=|\phi_{\phi\cdot\tilde{c},\phi^{\prime}\cdot\phi\cdot\tilde{c}}%
\phi_{\tilde{c},\phi\cdot\tilde{c}}\cdot c\rangle=|\phi_{\tilde{c}%
,(\phi^{\prime}\circ\phi)\cdot\tilde{c}}\cdot c\rangle=\hat{U}(\phi^{\prime
}\circ\phi)|c\rangle
\end{equation}
The action of the $\hat{U}(\phi)$ on $\mathcal{V}_{\mathrm{LMI}}$ descends
from the action on $\mathcal{H}_{\mathrm{kin}}$ via%
\begin{equation}
\langle c|\hat{U}(\phi)^{\prime}:=(\hat{U}(\phi^{-1})|c\rangle)^{\dag}%
=|\phi_{\tilde{c}(c),\phi^{-1}\cdot\tilde{c}(c)}\cdot c\rangle^{\dag}%
=\langle\phi_{\tilde{c}(c),\phi^{-1}\cdot\tilde{c}(c)}\cdot c|
\label{eq:novel-diff1}%
\end{equation}
It is easy to see that using above the equation we have,
\begin{equation}
(\Psi_{\lbrack\tilde{c}]_{1}}^{f^{1}}|\hat{U}(\phi)^{\prime}=\sum
_{(c_{1}^{\prime},c_{2},c_{3})\in\lbrack\tilde{c}]_{1}}f^{1}(\bar{V}%
(c_{1}^{\prime}\cup c_{2}\cup c_{3}))\langle\phi_{\tilde{c},\phi^{-1}%
\cdot\tilde{c}}\cdot(c_{1}^{\prime},c_{2},c_{3})|.
\end{equation}
where by $\tilde{c}$ we mean the WEO vertex-free charge network $\tilde
{c}(c_{1}^{\prime},\tilde{c}_{2},\tilde{c}_{3})$ underlying $(c_{1}^{\prime
},\tilde{c}_{2},\tilde{c}_{3})$ which is exactly $\tilde{c}$ for all
$(c_{1}^{\prime},\tilde{c}_{2},\tilde{c}_{3})\in\lbrack\tilde{c}]$.\newline
Now note that the vertex set $\bar{V}(c_{1}^{\prime}\cup c_{2}\cup c_{3})$
transforms \emph{equivariantly} under diffeomorphisms:%
\begin{equation}
\phi_{\tilde{c},\phi\tilde{c}}(\bar{V}(c_{1}^{\prime}\cup\tilde{c}_{2}%
\cup\tilde{c}_{3}))=\bar{V}(\phi_{\tilde{c},\phi\tilde{c}}(c_{1}^{\prime}%
)\cup\phi_{\tilde{c},\phi\tilde{c}}\tilde{c}_{2}\cup\phi_{\tilde{c},\phi
\tilde{c}}\tilde{c}_{3})
\end{equation}
We can now define a (dual) representation on $\mathcal{V}_{\mathrm{LMI}}$
based on the representation given in (\ref{eq:novel-diff1}) as follows:
\begin{equation}
\hat{U}(\phi)^{\prime}\Psi_{\lbrack\tilde{c}]_{i}}^{f^{i}}:=\Psi_{\phi
_{\tilde{c},\phi^{-1}\cdot\tilde{c}}\cdot\lbrack\tilde{c}]_{i}}^{f^{i}%
\circ\phi_{\phi^{-1}\cdot\tilde{c},\tilde{c}}} \label{novel-diff2}%
\end{equation}
$\forall\ i$. The above definition is justified by the following lemma.

\textit{Lemma}: $\Psi_{\lbrack\tilde{c}]_{i}}^{f^{i}}\left(  \hat{U}%
(\phi)|c_{A}\rangle\right)  =\Psi_{\phi_{\tilde{c},\phi^{-1}\tilde{c}}%
\cdot\lbrack\tilde{c}]_{i}}^{f^{i}\circ\phi_{\phi^{-1}\cdot\tilde{c},\tilde
{c}}}|c_{A}\rangle$ for all $|c_{A}\rangle\in\mathcal{H}_{\mathrm{kin}}$.

\textit{Proof}: We give the proof for $i=1$. Without loss of generality, we
assume that all WEO vertices of $c_{A}$ are of type 1, as otherwise both sides
are trivially zero.

Compute%
\begin{equation}
\Psi_{\lbrack\tilde{c}]_{1}}^{f^{1}}\left(  \hat{U}(\phi)|c_{A}\rangle\right)
=\sum_{c^{\prime}\in\lbrack\tilde{c}]_{1}}f^{1}\left(  \bar{V}(c^{\prime
})\right)  \delta_{c^{\prime},\phi_{\tilde{c}_{A},\phi\cdot\tilde{c}_{A}}}
\label{noveldiffonLMI-1}%
\end{equation}
where $c^{\prime}=(c_{1}^{\prime},\tilde{c}_{2},\tilde{c}_{3})\in\lbrack
\tilde{c}]_{1}$, and $\tilde{c}_{A}$ is the WEO vertex-free charge network
associated to $c_{A}$. We have that
\begin{equation}
\phi_{\tilde{c}_{A},\phi\cdot\tilde{c}_{A}}\cdot c_{A}\notin\lbrack\tilde
{c}_{1}]_{1}\qquad\Leftrightarrow\qquad\phi\cdot\tilde{c}_{A}\neq\tilde{c},
\end{equation}
and in this case, $\Psi_{\lbrack\tilde{c}]_{1}}^{f^{1}}\left(  \hat{U}%
(\phi)|c_{A}\rangle\right)  =0.$ On the other hand, if $\phi_{\tilde{c}%
_{A},\phi\cdot\tilde{c}_{A}}\cdot c_{A}\in\lbrack\tilde{c}]_{1}\ $%
($\Leftrightarrow\phi\cdot\tilde{c}_{A}=\tilde{c}$), then
\begin{equation}
\Psi_{\lbrack\tilde{c}]_{1}}^{f^{1}}\left(  \hat{U}(\phi)|c_{A}\rangle\right)
=f^{1}\left(  \bar{V}(\phi_{\tilde{c}_{A},\phi\cdot\tilde{c}_{A}}\cdot
c_{A})\right)  =f^{1}\left(  \bar{V}(\phi_{\phi^{-1}\cdot\tilde{c},\tilde{c}%
}\cdot c_{A})\right)  , \label{noveldiffonLMI-2}%
\end{equation}
whereas the RHS is given by
\begin{equation}
\sum_{c^{\prime}\in\lbrack\tilde{c}]_{1}}f^{1}\left(  \bar{V}(c^{\prime
})\right)  \delta_{c^{\prime},\phi_{\tilde{c}_{A},\phi\cdot\tilde{c}_{A}}%
}=\sum_{c^{\prime}\in\phi_{\tilde{c},\phi^{-1}\cdot\tilde{c}}\cdot
\lbrack\tilde{c}]_{1}}f^{1}\circ\phi_{\phi^{-1}\cdot\tilde{c},\tilde{c}%
}\left(  \bar{V}(c^{\prime})\right)  \delta_{c^{\prime},c_{A}}%
\end{equation}
Whence if $c_{A}\notin\phi_{\tilde{c},\phi^{-1}\cdot\tilde{c}}\cdot
\lbrack\tilde{c}]_{1}$ ($\Leftrightarrow\tilde{c}_{A}\neq\phi^{-1}\cdot
\tilde{c}$), then the right hand side of (\ref{noveldiffonLMI-1}) vanishes. On
the other hand if $c_{A}\in\phi_{\tilde{c},\phi^{-1}\cdot\tilde{c}}%
\cdot\lbrack\tilde{c}]_{1}\ $($\Leftrightarrow\tilde{c}_{A}=\phi^{-1}%
\cdot\tilde{c}$), then
\begin{equation}
\text{RHS}=f^{1}\circ\phi_{\phi^{-1}\cdot\tilde{c},\tilde{c}}\left(  \bar
{V}(c_{A})\right)  =f^{1}\left(  \bar{V}((\phi_{\phi^{-1}\cdot\tilde{c}%
,\tilde{c}}\cdot c_{A})\right)
\end{equation}
In the second equality we have used the fact that $\bar{V}$ is a
diffeomorphism-equivariant map on the set of vertices. This proves the lemma.

It is straightforward to verify that (\ref{novel-diff2}) defines a
representation:%
\begin{equation}
\hat{U}(\phi)^{\prime}\Psi_{\lbrack\tilde{c}]_{i}}^{f^{i}}:=\Psi_{\phi
_{\tilde{c},\phi^{-1}\cdot\tilde{c}}\cdot\lbrack\tilde{c}]_{i}}^{f^{i}%
\circ\phi_{\phi^{-1}\cdot\tilde{c},\tilde{c}}}%
\end{equation}%
\[
\hat{U}(\phi_{1})^{\prime}\hat{U}(\phi_{2})^{\prime}\Psi_{\lbrack\tilde
{c}]_{1}}^{f^{1}}=\hat{U}(\phi_{1})^{\prime}\Psi_{\phi_{\tilde{c},\phi
_{2}^{-1}\cdot\tilde{c}}\cdot\lbrack\tilde{c}]_{i}}^{f^{i}\circ\phi_{\phi
_{2}^{-1}\cdot\tilde{c},\tilde{c}}}%
\]%
\begin{equation}
\hat{U}(\phi_{1})^{\prime}\hat{U}(\phi_{2})^{\prime}\Psi_{\lbrack\tilde
{c}]_{1}}^{f^{1}}=(\Psi_{\lbrack\phi_{2}^{-1}\phi_{1}^{-1}\tilde{c}]_{1}%
}^{f^{1}\circ\alpha_{\tilde{c},\phi_{1}^{-1}\tilde{c}}^{-1}\circ\alpha
_{\phi_{1}^{-1}\tilde{c},\phi_{2}^{-1}\phi_{1}^{-1}\tilde{c}}^{-1}}%
|~=(\Psi_{\lbrack(\phi_{1}\circ\phi_{2})^{-1}\tilde{c}]_{1}}^{f^{1}\circ
\alpha_{\tilde{c},(\phi_{1}\circ\phi_{2})^{-1}\tilde{c}}^{-1}}|~=(\Psi
_{\lbrack\tilde{c}]_{1}}^{f^{1}}|\hat{U}(\phi_{1}\circ\phi_{2})^{\prime},
\label{indeedarep}%
\end{equation}
so that we indeed have a representation.

\bigskip
We now point out a rather interesting property of this representation. Although ${\cal V}_{LMI}$ is a subspace of distributions on ${\cal H}_{kin}$ , there is a canonical choice of inner product on this space.\footnote{Note that on the Lewandowski Marolf Habitat defined in \cite{lm} no such canonical choice exists!}\\
Given $\Psi_{\lbrack\tilde{c}^{\prime}]_{i}}^{f_{1}^{i}},\Psi
_{\lbrack\tilde{c}^{\prime\prime}]_{j}}^{f_{2}^{j}},$ 
the inner product is defined as,
\begin{equation}\label{innerproduct}
\begin{array}{lll}
\left(\Psi_{\lbrack\tilde{c}^{\prime}]_{i}}^{f_{1}^{i}},\Psi_{\lbrack\tilde{c}^{\prime\prime}]_{j}%
}^{f_{2}^{j}}\right):=\delta_{\tilde{c}^{\prime},\tilde{c}^{\prime\prime}}\bar{f}_{1}^{i}%
(V(\tilde{c}^{\prime}))^{*}f_{2}^{j}(V(\tilde{c}^{\prime\prime}))\delta_{i,j}
\end{array}
\end{equation}
Note that this inner product is not positive definite unless we restrict ourselves to ``basis" states $\Psi^{f^{i}}_{[\tilde{c}]_{i}}$ in ${\cal V}_{LMI}$ which are such that $\vert f^{i}(V(\tilde{c}))\vert^{2}\ >\ 0$.\\
In any case, it is clear that the representation of $Diff(\Sigma)$ on ${\cal V}_{LMI}$ is unitary with respect to (\ref{innerproduct}) as,
\begin{equation}
\begin{array}{lll}
\left(\hat{U}(\phi)\Psi_{\lbrack\tilde{c}^{\prime}]_{i}}^{f_{1}^{i}},\hat{U}(\phi)\Psi_{\lbrack\tilde{c}^{\prime\prime}]_{j}%
}^{f_{2}^{j}}\right)=\\
\vspace*{0.1in}
\left(\Psi_{\phi_{\tilde{c}^{\prime},\phi^{-1}\cdot\tilde{c}^{\prime}}\lbrack\tilde{c}^{\prime}]_{i}}^{f_{1}^{i}\circ\phi_{\phi^{-1}\tilde{c}^{\prime},\tilde{c}^{\prime}}},\Psi_{\phi_{\tilde{c}^{\prime\prime},\phi^{-1}\cdot\tilde{c}^{\prime\prime}}\lbrack\tilde{c}^{\prime\prime}]_{j}%
}^{f_{2}^{j}\circ\phi_{\phi^{-1}\tilde{c}^{\prime\prime},\cdot\tilde{c}^{\prime\prime}}}\right)=\\
\vspace*{0.1in}
\delta_{\phi\cdot \tilde{c}^{\prime},\phi\cdot \tilde{c}^{\prime\prime}}\\
\hspace*{0.3in}(f_{1}^{i}\circ\phi_{\phi^{-1}\tilde{c}^{\prime},\tilde{c}^{\prime}})(V(\phi^{-1}\tilde{c}^{\prime}))^{*}\ (f_{2}^{j}\circ\phi_{\phi^{-1}\tilde{c}^{\prime\prime},\cdot\tilde{c}^{\prime\prime}})(V(\phi^{-1}\tilde{c}^{\prime\prime}))\ \delta_{i,j}\\
\vspace*{0.1in}
=\ \delta_{\phi\cdot \tilde{c}^{\prime},\phi\cdot \tilde{c}^{\prime\prime}}
f_{1}^{i}(V(\tilde{c}^{\prime}))^{*}f_{2}^{j}(V(\tilde{c}^{\prime\prime}))\delta_{i,j}\\
\vspace*{0.1in}
=\ \left(\Psi_{\lbrack\tilde{c}^{\prime}]_{i}}^{f_{1}^{i}},\Psi_{\lbrack\tilde{c}^{\prime\prime}]_{j}%
}^{f_{2}^{j}}\right)
\end{array}
\end{equation}
where we have used the fact that $\phi_{\tilde{c},\phi\cdot\tilde{c}}\vert_{\tilde{c}}\ =\ \phi\vert_{\tilde{c}}$ $\forall\ \phi$.

\bigskip

\bigskip

\section{Diffeomorphism Covariance}

\label{covariance}

\subsection{$\hat{H}^{1}[N]^{\prime}\Psi_{\lbrack c]_{1}}^{f^{1}}$}\label{H1onf1}

In this section we revisit the diffeomorphism-covariance of $\hat{H}%
^{1}[N]^{\prime}$ on $\Psi_{\lbrack\tilde{c}]_{1}}^{f^{1}}$ in light of the
new representation of Diff($\Sigma$) on $\mathcal{V}_{LMI}$ involving
$\phi_{\tilde{c}_{0},\tilde{c}}$ maps.\newline Whence our aim is to check if
\begin{equation}
\left(  \hat{U}(\phi)^{\prime}\hat{H}^{1}[N]^{\prime}\hat{U}(\phi
^{-1})^{\prime}\Psi_{\lbrack\tilde{c}]_{1}}^{f^{1}}\right)  |c_{A}%
\rangle\ =\ \left(  \hat{H}^{1}[(\phi_{\tilde{c},\phi\cdot\tilde{c}})^{\ast
}N]^{\prime}\Psi_{\lbrack\tilde{c}]_{1}}^{f^{1}}\right)  |c_{A}\rangle
\label{diffcovforH1onLMI1}%
\end{equation}
$\forall\ \Psi_{\lbrack\tilde{c}]_{1}}^{f^{1}}\ \in\mathcal{V}_{LMI}$ and
$|c_{A}\rangle\ \in\mathcal{H}_{kin}$.\newline Note that on the right hand
side, we expect the lapse to be pulled back by the diffeomorphism
$\phi_{\tilde{c},\phi\cdot\tilde{c}}$ (which given a $\tilde{c}$ and a $\phi$
is fixed once and for all) and not by $\phi$ as warranted by the new
representation of the Diffeomorphism group.\newline Before proceeding with the
computation, we outline our setup which will also help us clarifying our
(often confusing) notations. We denote the reference charge-network in
$[\tilde{c}]_{diff}$ by $\tilde{c}^{0}$. The WEO vertex free state underlying
$c_{A}$ will be denoted by $\tilde{c}_{A}$. Given a vertex $v_{A}$ in
$\tilde{c}$, we will denote the corresponding (image under diffeomorphism
$\phi_{\tilde{c},\tilde{c}^{0}}$) vertex in $\tilde{c}^{0}$ as $v_{A}^{0}%
$.\newline

Without loss of generality we assume that the only WEO vertices which belong
to $V(c_{A})$ are of $type-1$, as otherwise both sides are trivially
zero.\newline We first compute the LHS using (\ref{eq:novel-diff1}),
(\ref{novel-diff2}) and (\ref{H1onLMI})
\begin{equation}%
\begin{array}
[c]{lll}%
\text{LHS}\ =\ \left(  \hat{U}(\phi)^{\prime}\hat{H}^{1}[N]^{\prime}\hat
{U}(\phi^{-1})^{\prime}\Psi^{f^{1}}_{[\tilde{c}]_{1}}\right)  \vert
c_{A}\rangle &  & \\
\vspace*{0.1in} \hspace*{0.5in}=\ \left(  \hat{H}^{1}[N]^{\prime}\hat{U}%
(\phi^{-1})^{\prime})\Psi^{f^{1}}_{[\tilde{c}]_{1}}\right)  \vert\phi
_{\tilde{c}_{A},\phi\cdot\tilde{c}_{A}}\cdot c_{A}\rangle &  & \\
\vspace*{0.1in} \hspace*{0.5in}=\ \left(  \hat{H}^{1}[N]^{\prime}\Psi
^{f^{1}\circ\phi_{\phi\cdot\tilde{c},\tilde{c}}}_{\phi_{\tilde{c},\phi
\cdot\tilde{c}}\cdot[\tilde{c}_{1}]}\right)  \vert\phi_{\tilde{c}_{A}%
,\phi\cdot\tilde{c}_{A}}\cdot c_{A}\rangle &  & \\
\vspace*{0.1in} \hspace*{0.5in}=\ \sum_{v\in V(\tilde{c})}\left(
\Psi^{(f\circ\phi_{\phi\cdot\tilde{c},\tilde{c}})^{1,1}_{v,2}[N]}%
_{\phi_{\tilde{c},\phi\cdot\tilde{c}}\cdot[\tilde{c}_{1}]}\ -\ \Psi
^{(f\circ\phi_{\phi\cdot\tilde{c},\tilde{c}})^{1,1}_{v,3}[N]}_{\phi_{\tilde
{c},\phi\cdot\tilde{c}}\cdot[\tilde{c}_{1}]}\right)  \vert\phi_{\tilde{c}%
_{A},\phi\cdot\tilde{c}_{A}}\cdot c_{A}\rangle &  &
\end{array}
\end{equation}

Further analysis of above equation can be divided into following two (mutually
exclusive and exhaustive) cases.\newline

\underline{\textbf{Case 1}} : $\phi_{\tilde{c}_{A},\phi\cdot\tilde{c}_{A}%
}\cdot c_{A}\ \notin\ \phi_{\tilde{c},\phi\cdot\tilde{c}}\cdot\lbrack\tilde
{c}_{1}]\ \Leftrightarrow\ \phi\cdot\tilde{c}_{A}\ \neq\ \phi\cdot\tilde
{c}\ \Leftrightarrow\ \tilde{c}_{A}\ \neq\ \tilde{c}$.\newline In this case it
is clear that LHS = 0.\newline\underline{\textbf{Case 2}} : $\phi_{\tilde
{c}_{A},\phi\cdot\tilde{c}_{A}}\cdot c_{A}\ \in\ \phi_{\tilde{c},\phi
\cdot\tilde{c}}\cdot\lbrack\tilde{c}_{1}]\ \Leftrightarrow\ \phi\cdot\tilde
{c}_{A}\ =\ \phi\cdot\tilde{c}\ \Leftrightarrow\ \tilde{c}_{A}\ =\ \tilde{c}%
$.\newline In this case,
\begin{equation}%
\begin{array}
[c]{lll}%
\boxed{\textrm{LHS}\ =\ \sum_{v\in V(\tilde{c})}\left(\left(f\circ\phi_{\phi\cdot\tilde{c},\tilde{c}}\right)^{1,1}_{v,2}[N]\left(\overline{V}(\phi_{\tilde{c}_{A},\phi\cdot\tilde{c}_{A}})\right)\ -\ \left(f\circ\phi_{\phi\cdot\tilde{c},\tilde{c}}\right)^{1,1}_{v,3}[N]\left(\overline{V}(\phi_{\tilde{c}_{A},\phi\cdot\tilde{c}_{A}})\right)\right)} &
&
\end{array}
\end{equation}
For the sake of pedagogy, we now assume that the only vertex in $V(\phi
\cdot\tilde{c})\ =\ V(\phi\cdot\tilde{c}_{A})$ which falls inside the support
of $N$ is a vertex $v_{A}$. As the Hamiltonian constraint action is linearly
distributed over vertices, there is no loss of generality in this
assumption.\newline In this case \textbf{Case 2} gets further sub-divided into
following two cases.\newline\underline{\textbf{case 2a}} : $v_{A}%
\ \notin\ \overline{V}(\phi_{\tilde{c}_{A},\phi\cdot\tilde{c}_{A}}\cdot
c_{A})$, and in this case,
\begin{equation}%
\begin{array}
[c]{lll}%
\text{LHS}\ =\ \sum_{v\in V(\tilde{c})}\left(  \left(  f\circ\phi_{\phi
\cdot\tilde{c},\tilde{c}}\right)  _{v,2}^{1,1}[N]\left(  \overline{V}%
(\phi_{\tilde{c}_{A},\phi\cdot\tilde{c}_{A}}\cdot c_{A})\right)  \ -\ \left(
f\circ\phi_{\phi\cdot\tilde{c},\tilde{c}}\right)  _{v,3}^{1,1}[N]\left(
\overline{V}(\phi_{\tilde{c}_{A},\phi\cdot\tilde{c}_{A}}\cdot c_{A})\right)
\right)  &  & \\
\vspace*{0.1in}\hspace*{0.6in}\ =\ \left(  \left(  f\circ\phi_{\phi\cdot
\tilde{c},\tilde{c}}\right)  _{v_{A},2}^{1,1}[N]\left(  \overline{V}%
(\phi_{\tilde{c}_{A},\phi\cdot\tilde{c}_{A}}\cdot v_{A})\right)  \ -\ \left(
f\circ\phi_{\phi\cdot\tilde{c},\tilde{c}}\right)  _{v_{A},3}^{1,1}[N]\left(
\overline{V}(\phi_{\tilde{c}_{A},\phi\cdot\tilde{c}_{A}}\cdot v_{A})\right)
\right)  &  & \\
\vspace*{0.1in}\hspace*{0.6in}=\ \left(  f\circ\phi_{\phi\cdot\tilde{c}%
,\tilde{c}}\right)  ^{1}\left(  \overline{V}(\phi_{\tilde{c}_{A},\phi
\cdot\tilde{c}_{A}}\cdot v_{A})\right)  \ -\ \left(  f\circ\phi_{\phi
\cdot\tilde{c},\tilde{c}}\right)  ^{1}\left(  \overline{V}(\phi_{\tilde{c}%
_{A},\phi\cdot\tilde{c}_{A}}\cdot c_{A})\right)  &  & \\
\vspace*{0.1in}\hspace*{0.6in}=\ 0 &  &
\end{array}
\label{lhscase2a}%
\end{equation}
where in the second line we have used the assumption stated above and in the
third line, we have used the defining property of $f^{1,1}$ functions,
\begin{equation}
f_{v,2}^{1,1}[N]\left(  \overline{V}(c^{\prime})\right)  \ =\ f^{1}\left(
\overline{V}(c^{\prime})\right)
\end{equation}
if $v\ \notin\ \overline{V}(c^{\prime})$.\newline The only case where LHS is
non-trivial is given by,\newline\underline{\textbf{case 2b}} : $v_{A}%
\in\ \overline{V}(\phi_{\tilde{c}_{A},\phi\cdot\tilde{c}_{A}}\cdot c_{A})$.
(Recall that $v_{A}\ \in\ V(\phi\cdot\tilde{c}_{A})$ by definition.)\newline
In this case, we can start off with the second line in (\ref{lhscase2a}) and
evaluate the LHS.
\begin{equation}%
\begin{array}
[c]{lll}%
\text{LHS}\ =\ \left(  \left(  f\circ\phi_{\phi\cdot\tilde{c},\tilde{c}%
}\right)  _{v_{A},2}^{1,1}[N]\left(  \overline{V}(\phi_{\tilde{c}_{A}%
,\phi\cdot\tilde{c}_{A}}\cdot v_{A})\right)  \ -\ \left(  f\circ\phi
_{\phi\cdot\tilde{c},\tilde{c}}\right)  _{v_{A},3}^{1,1}[N]\left(
\overline{V}(\phi_{\tilde{c}_{A},\phi\cdot\tilde{c}_{A}}\cdot v_{A})\right)
\right)  &  & \\
\vspace*{0.1in}\hspace*{0.3in}=\ \left(  f\circ\phi_{\phi\cdot\tilde{c}%
,\tilde{c}}\right)  _{v_{A},2}^{1,1}[N]\left(  v_{A},\overline{V}%
((\phi_{\tilde{c}_{A},\phi\cdot\tilde{c}_{A}}\cdot v_{A})-\{v_{A}\}\right)
\ -\ \left(  f\circ\phi_{\phi\cdot\tilde{c},\tilde{c}}\right)  _{v_{A}%
,3}^{1,1}[N]\left(  v_{A},\overline{V}((\phi_{\tilde{c}_{A},\phi\cdot\tilde
{c}_{A}}\cdot v_{A})-\{v_{A}\}\right)  &  & \\
&  &
\end{array}
\end{equation}
Here without loss of generality we have assumed that the first argument of
$f^{1}\circ\phi_{\phi\cdot\tilde{c},\tilde{c}}$ is $v_{A}$.\newline Notice
that as $\phi_{\tilde{c}_{A},\phi\cdot\tilde{c}_{A}}\cdot c_{A}\ \in
\ \phi_{\tilde{c},\phi\cdot\tilde{c}}\cdot\lbrack\tilde{c}]_{1}$ , the Lemma
(A.1) in the appendix tells us that $\tilde{c}_{A}\ =\ \tilde{c}$. We can now use
(\ref{functionforH1}) in the above equation along with the fact that
$\tilde{c}_{A}\ =\ \tilde{c}$ and get
\begin{equation}%
\begin{array}
[c]{lll}%
\text{LHS}\ =\ N(v_{A},\{x_{v_{A}}\}^{\phi\cdot\tilde{c}})\lambda(\vec
{n}_{v_{A}}^{\phi\cdot\tilde{c}})\left[  V_{2}^{a}(v_{A},\tilde{c})-V_{3}%
^{a}(v_{A},\tilde{c})\right]  \frac{\partial}{\partial(x_{v_{A}}^{\phi
\cdot\tilde{c}})^{a}}\left(  \phi_{\phi\cdot\tilde{c},\tilde{c}}^{\ast
}f\right)  (v_{A},\{\text{.,.,.}\}) &  & \\
&  &
\end{array}
\end{equation}
Recall that the components of quantum-shift $V_{i}^{a}(v_{A},\tilde{c})$ are
evaluated in the co-ordinate chart $\{x_{v_{A}}\}^{\phi\cdot\tilde{c}}$ which
is centered at $v_{A}$ and is obtained by the push-forward of $\{x_{v_{A}^{0}%
}\}^{\tilde{c}^{0}}$ centered at a vertex $v_{A}^{0}\ \in\ V(\tilde{c}^{0})$.
Thus components of $\vec{V}(v_{A},\tilde{c})$ in $\{x_{v_{A}}\}^{\tilde{c}}$
are equal to the components of $\vec{V}(v_{A}^{0},\tilde{c}^{0})$ in the
co-ordinate system $\{x_{v_{A}^{0}}\}^{\tilde{c}^{0}}$. Using this, above
equation simplifies to,
\begin{equation}%
\begin{array}
[c]{lll}%
\text{LHS}=\ N(v_{A},\{x_{v_{A}}\}^{\phi\cdot\tilde{c}})\lambda(\vec{n}%
_{v_{A}^{0}}^{\tilde{c}^{0}})\left[  V_{2}^{a^{\prime}}(v_{A}^{0},\tilde
{c}^{0})\ -\ V_{3}^{a^{\prime}}(v_{A}^{0},\tilde{c}^{0})\right]
\frac{\partial}{\partial(x_{v_{A}^{0}}^{\phi\cdot\tilde{c}})^{a^{\prime}}%
}\left(  \phi_{c^{0},\phi\cdot\tilde{c}}^{\ast}\circ\phi_{\phi\cdot\tilde
{c},\tilde{c}}^{\ast}f\right)  (v_{A}^{0},\{\text{.,.,.}\}) &  & \\
\vspace*{0.1in}%
\boxed{\textrm{LHS}=\ N(v_{A},\{x_{v_{A}}\}^{\phi\cdot\tilde{c}})\lambda(\vec{n}_{v_{A}^{0}}^{\tilde{c}^{0}})\left[V_{2}^{a^{\prime}}(v_{A}^{0},\tilde{c}^{0})\ -\ V_{3}^{a^{\prime}}(v_{A}^{0},\tilde{c}^{0})\right]\frac{\partial}{\partial(x_{v_{A}^{0}}^{\phi\cdot\tilde{c}})^{a^{\prime}}}\left(\phi^{*}_{c^{0},\tilde{c}}f\right)(v_{A}^{0},\{\textrm{.,.,.}\})} &
&
\end{array}
\label{case2bLHSfinal}%
\end{equation}

In the above equations we have also explicitly displayed the dependence of
density-weighted lapse on co-ordinate system.\newline We now evaluate the RHS
in (\ref{diffcovforH1onLMI1})
\begin{equation}
\text{RHS}\ =\ \left(  \hat{H}^{1}[(\phi_{\tilde{c},\phi\cdot\tilde{c}}%
)^{*}N]\Psi^{f^{1}}_{[\tilde{c}]_{1}}\right)  \vert c_{A}\rangle
\end{equation}
As the only vertex in $V(\phi\cdot\tilde{c}_{A})$ which is inside the support
of $N$ is $v_{A}$, it implies that the only vertex in $V(\tilde{c}_{A})$ which
falls inside the support of $(\phi_{\tilde{c},\phi\cdot\tilde{c}})^{*}N$ is
$\phi^{-1}\cdot v_{A}$\newline As before we analyze two cases (\textbf{Case
1}) and (\textbf{Case 2}) separately.\newline\underline{\textbf{Case 1}}
:\newline Recall that case-1 corresponds to $\tilde{c}_{A}\ \neq\ \tilde{c}$
in which case it is easy to see that
\begin{equation}%
\begin{array}
[c]{lll}%
\text{RHS}\ =\ \sum_{v\ \in\ V(\tilde{c})}\left(  \Psi^{f^{1,2}_{v}[\phi
^{*}N]}_{[\tilde{c}]_{1}}\ -\ \Psi^{f^{1,3}_{v}[\Phi^{*}N]}_{[\tilde{c}]_{1}%
}\right)  \vert c_{A}\rangle\ =\ 0\ =\ \text{LHS} &  &
\end{array}
\end{equation}
\underline{\textbf{Case 2}} :\newline This is the complementary case where
$\tilde{c}_{A}\ =\ \tilde{c}$.\newline While analyzing LHS in case-2, we
specialized to the situation where the only vertex in $V(\phi\cdot\tilde
{c})\ =\ V(\phi\cdot\tilde{c}_{A})$ which is inside the support of $N$ is
$v_{A}$. Clearly this implies that the only vertex in $V(\tilde{c}%
_{A})\ =\ V(\tilde{c})$ which lies in the support of $\phi^{*}N$ is $\phi
^{-1}\cdot v_{A}$. In this case, RHS is given by,
\begin{equation}%
\begin{array}
[c]{lll}%
\text{RHS}\ =\ \left(  f^{1,2}_{\phi^{-1}\cdot v_{A}}\left(  \overline
{V}(c_{A})\right)  \ -\ f^{1,3}_{\phi^{-1}\cdot v_{A}}\left(  \overline
{V}(c_{A})\right)  \right)  &  &
\end{array}
\end{equation}

As in the case of evaluation of LHS, this case can be further analyzed by
looking at to sub-cases (\textbf{case-2a}) and (\textbf{case-2b})
separately.\newline\underline{\textbf{case 2a}} : $v_{A}\ \notin\ \overline
{V}(\phi_{\tilde{c}_{A},\phi\cdot\tilde{c}_{A}}\cdot c_{A})$.\newline As
$\overline{V}$ is a diffeomorphism equivariant map, we have
\[%
\begin{array}
[c]{lll}%
\phi_{\phi\cdot\tilde{c}_{A},\tilde{c}_{A}}\cdot v_{A}\ \notin\ \overline
{V}(c_{A}) &  & \\
\vspace*{0.1in} \implies\phi^{-1}\cdot v_{A}\ \notin\ \overline{V}(c_{A}) &
&
\end{array}
\]
The second line in the above equation needs and explaination.\newline As
$v_{A}\ \in V(\phi_{\tilde{c}_{A},\phi\cdot\tilde{c}_{A}}\cdot c_{A})$ but
$v_{A}\ \notin\ \overline{V}(\phi_{\tilde{c}_{A},\phi\cdot\tilde{c}_{A}}\cdot
c_{A})$, it is clear that $v_{A}\ \in\ V(\phi\cdot\tilde{c}_{A})$. But on
$V(\phi\cdot\tilde{c}_{A})$, $\phi_{\phi\cdot\tilde{c}_{A},\tilde{c}_{A}%
}\ =\ \phi^{-1}$, which is used in the second line of the above
equation.\newline However if $\phi^{-1}\cdot v_{A}\ \notin\ \overline{V}%
(c_{A})$ we have
\begin{equation}
\label{case2aRHS}%
\begin{array}
[c]{lll}%
\text{RHS}\ =\ \left(  f^{1,2}_{\phi^{-1}\cdot v_{A}}\left(  \overline
{V}(c_{A})\right)  \ -\ f^{1,3}_{\phi^{-1}\cdot v_{A}}\left(  \overline
{V}(c_{A})\right)  \right)  &  & \\
\vspace*{0.1in} \hspace*{0.5in}=\ \left(  f\left(  \overline{V}(c_{A})\right)
\ -\ f\left(  \overline{V}(c_{A})\right)  \right)  \ =\ 0 &  &
\end{array}
\end{equation}
Whence even in this case we get
\begin{equation}
\text{LHS}\ =\ \text{RHS}\nonumber
\end{equation}
we are finally left with the final and only non-trivial case \textbf{case-2b}%
.\newline\underline{\textbf{Case 2b}} : $v_{A}\ \in\ \overline{V}(\phi
_{\tilde{c}_{A},\phi\cdot\tilde{c}_{A}}\cdot c_{A})$.\newline An argument
similar to the one given above (\ref{case2aRHS}) leads us to conclude that
$\phi^{-1}\cdot v_{A}\ \in\ \overline{V}(c_{A})$. Whence in this case, RHS is
given by,
\begin{equation}%
\begin{array}
[c]{lll}%
\text{RHS}\ =\ \left(  f^{1,2}_{\phi^{-1}\cdot v_{A}}\left(  \overline
{V}(c_{A})\right)  \ -\ f^{1,3}_{\phi^{-1}\cdot v_{A}}\left(  \overline
{V}(c_{A})\right)  \right)  &  & \\
\vspace*{0.1in} \hspace*{0.5in}=\ \left(  f^{1,2}_{\phi^{-1}\cdot v_{A}%
}\left(  \phi^{-1}\cdot v_{A}, \overline{V}(c_{A})-\{\phi^{-1}\cdot
v_{A}\}\right)  \ -\ f^{1,3}_{\phi^{-1}\cdot v_{A}}\left(  \phi^{-1}\cdot
v_{A}, \overline{V}(c_{A})-\{\phi^{-1}\cdot v_{A}\}\right)  \right)  &  & \\
\vspace*{0.1in}
\boxed{\textrm{RHS}\ =(\phi_{\tilde{c},\phi\cdot\tilde{c}})^{*}N(\phi^{-1}v_{A}, \{x_{\phi^{-1}\cdot v_{A}}\}^{\tilde{c}_{A}})\lambda(\vec{n}_{\phi^{-1}v_{A}}^{\tilde{c}_{A}})\left[V_{2}^{a}(\phi^{-1}v_{A},\tilde{c})-V_{3}^{a}(\phi^{-1}v_{A},\tilde{c})\right] \left(\frac{\partial}{\partial (x_{\phi^{-1}\cdot v_{A}}^{\tilde{c}_{A}})^{a}}f\right)(\phi^{-1}v_{A},\textrm{.,.,.})} &
&
\end{array}
\end{equation}
Once again (in exact analogy with the way we arrived at (\ref{case2bLHSfinal}%
)) we can use the following three observations to ``pull back" the above
equation to $(v_{A}^{0},\tilde{c}^{0})$.\newline\noindent{(1)} $\vec{V}%
_{2}(\phi^{-1}\cdot v_{A},\tilde{c})$ is obtained by push-forward of $\vec
{V}_{2}(v_{A}^{0},\tilde{c}^{0})$ using $(\phi_{\tilde{c}^{0},\tilde{c}})_{*}%
$, it implies that the (ordered set of) components $V^{a^{\prime\prime}}%
(\phi^{-1}v_{A},\tilde{c})$ in the preferred co-ordinate system $\{x_{\phi
^{-1}\cdot v_{A}}\}^{\tilde{c}}\ :=\ (\phi_{\tilde{c}^{0},\tilde{c}}%
)_{*}\{x_{v_{A}^{0}}\}^{\tilde{c}^{0}}$ centered at $\phi^{-1}\cdot v_{A}$ are
same as the components $V^{a}(v_{A}^{0},\tilde{c}^{0})$ of $\vec{V}(v_{A}%
^{0},\tilde{c}^{0})$ in the co-ordinate system $\{x_{v_{A}^{0}}\}^{\tilde
{c}^{0}}$.\newline\noindent{(2)} We also have, by construction $\lambda
(\vec{n}_{\phi^{-1}v_{A}}^{\tilde{c}_{A}})\ =\ \lambda(\vec{n}_{v_{A}^{0}%
}^{\tilde{c}^{0}})$.\newline\noindent{(3)}
\begin{equation}%
\begin{array}
[c]{lll}%
\frac{\partial}{\partial(x_{\phi^{-1}\cdot v_{A}}^{\tilde{c}})^{a}}%
\ f(\phi^{-1}v_{A},\text{.,.,.})= &  & \\
\vspace*{0.1in} \hspace*{0.3in}=\left(  (\phi_{\tilde{c}^{0},\tilde{c}}%
)_{*}\frac{\partial}{\partial(x_{v_{A}^{0}}^{\tilde{c}^{0}})^{a}}\right)
f(\phi^{-1}v_{A},\text{.,.,.}) &  & \\
\vspace*{0.1in} \hspace*{0.3in}=\ \frac{\partial}{\partial(x_{v_{A}^{0}%
}^{\tilde{c}^{0}})^{a}}\left(  \phi^{*}_{\tilde{c}^{0},\tilde{c}}f\right)
(v_{A}^{0},\text{.,.,.})\ \forall\ a &  &
\end{array}
\end{equation}
Whence
\begin{equation}
\label{case2bRHSfinal}%
\begin{array}
[c]{lll}%
\boxed{\textrm{RHS}\ =(\phi_{\tilde{c},\phi\cdot\tilde{c}})^{*}N(\phi^{-1}v_{A}, \{x_{\phi^{-1}\cdot v_{A}}\}^{\tilde{c}_{A}})\lambda(\vec{n}_{v_{A}^{0}}^{\tilde{c}^{0}})\left[V_{2}^{a}(v_{A}^{0},\tilde{c}^{0})-V_{3}^{a}(v_{A}^{0},\tilde{c}^{0})\right] \left(\frac{\partial}{\partial (x_{v_{A}^{0}}^{\tilde{c}^{0}})^{a}}\ \phi^{*}_{\tilde{c}^{0},\tilde{c}}f\right)(v_{A}^{0},\textrm{.,.,.})} &
&
\end{array}
\end{equation}
We can now compare the above equation with (\ref{case2bLHSfinal}) and see that
the only possible source of mismatch arises from the evaluation of Lapse. The
dependence of lapse in (\ref{case2bLHSfinal}) and (\ref{case2bRHSfinal}) are
respectively given by
\begin{equation}
\nonumber\\%
\begin{array}
[c]{lll}%
N(v_{A},\{x_{v_{A}}\}^{\phi\cdot\tilde{c}})\ =\ N\left(  v_{A},(\phi
_{\tilde{c}^{0},\phi\cdot\tilde{c}})_{*})\{x_{v_{A}^{0}}\}^{\tilde{c}^{0}%
}\right)  &  & \\
\vspace*{0.1in} (\phi_{\tilde{c},\phi\cdot\tilde{c}})^{*}N(\phi^{-1}v_{A},
\{x_{\phi^{-1}\cdot v_{A}}\}^{\tilde{c}_{A}})\ =\ N(v_{A},(\phi_{\tilde
{c},\phi\cdot\tilde{c}})_{*} \{x_{\phi^{-1}\cdot v_{A}}\}^{\tilde{c}_{A}%
})\ =\ N(v_{A},\phi_{*}(\phi_{\tilde{c}^{0},\tilde{c}})_{*} \{x_{v_{A}^{0}%
}\}^{\tilde{c}^{0}})\ =\ N\left(  v_{A},(\phi_{\tilde{c}^{0},\phi\cdot
\tilde{c}})_{*})\{x_{v_{A}^{0}}\}^{\tilde{c}^{0}}\right)  &  &
\end{array}
\end{equation}
Thus even for \textbf{Case-2b} we see that LHS equals the RHS.\newline Whence
we conclude that
\begin{equation}\label{covofH1onS1-final}
\begin{array}
[c]{lll}%
\boxed{\left(\hat{U}(\phi)\hat{H}^{1}[N]^{\prime}\hat{U}(\phi^{-1})\right)\Psi^{f^{1}}_{[\tilde{c}]_{1}}\ =\ \hat{H}^{1}[(\phi_{\tilde{c},\phi\cdot\tilde{c}})^{*}N]\Psi^{f^{1}}_{[\tilde{c}]_{1}}} &
&
\end{array}
\end{equation}

\subsection{$\hat{H}^{2}[N]\Psi_{\lbrack c]_{1}}^{f^{1}}$}\label{H2onf1}

In this section we will like to show that
\begin{equation}
\left(  \hat{U}(\phi)^{\prime}\hat{H}^{(2)}[N]^{\prime}\hat{U}(\phi
^{-1})^{\prime}\Psi_{\lbrack\tilde{c}^{1}]_{1}}^{f^{1}}\right)  (|c_{A}%
\rangle)=\left(  \hat{H}^{(2)}[(\phi_{\tilde{c},\phi\cdot\tilde{c}})^{\ast
}N]^{\prime}\Psi_{\lbrack\tilde{c}^{1}]_{1}}^{f^{1}}\right)  (|c_{A}\rangle)
\label{diffcovofH2onstate1}%
\end{equation}
$\forall\ \vert c_{A}\rangle\ \in\ \mathcal{H}_{kin}$, $\forall\ \phi
\ \in\ Diff(\Sigma)$ and $\forall\ N$.\newline

Once again without loss of generality we assume that the only WEO vertices
which belong to $V(c_{A})$ are of \textquotedblleft type-1" (i.e. all the
edges incident on any WEO vertex is only charged under $U(1)_{1}$), as
otherwise both sides are trivially zero.\newline We first compute the LHS
using (\ref{eq:novel-diff1}), (\ref{novel-diff2}) and (\ref{h2psi1})
\begin{equation}%
\begin{array}
[c]{lll}%
\text{LHS}\ =\ \left(  \hat{U}(\phi)^{\prime}\hat{H}^{2}[N]^{\prime}\hat
{U}(\phi^{-1})^{\prime}\Psi_{\lbrack\tilde{c}]_{1}}^{f^{1}}\right)
|c_{A}\rangle &  & \\
\vspace*{0.1in}\hspace*{0.5in}=\ \left(  \hat{H}^{2}[N]^{\prime}\hat{U}%
(\phi^{-1})^{\prime})\Psi_{\lbrack\tilde{c}]_{1}}^{f^{1}}\right)
|\phi_{\tilde{c}_{A},\phi\cdot\tilde{c}_{A}}\cdot c_{A}\rangle &  & \\
\vspace*{0.1in}\hspace*{0.5in}=\ \left(  \hat{H}^{2}[N]^{\prime}\Psi
_{\phi_{\tilde{c},\phi\cdot\tilde{c}}\cdot\lbrack\tilde{c}_{1}]}^{f^{1}%
\circ\phi_{\phi\cdot\tilde{c},\tilde{c}}}\right)  |\phi_{\tilde{c}_{A}%
,\phi\cdot\tilde{c}_{A}}\cdot c_{A}\rangle &  & \\
\vspace*{0.1in}\hspace*{0.5in}=\ \sum_{v\in V(\tilde{c})}\left(  \Psi
_{\phi_{\tilde{c},\phi\cdot\tilde{c}}\cdot\lbrack\tilde{c}_{1}]}^{(f\circ
\phi_{\phi\cdot\tilde{c},\tilde{c}})_{v,1}^{1,2}[N]}\ -\ \Psi_{\phi_{\tilde
{c},\phi\cdot\tilde{c}}\cdot\lbrack\tilde{c}_{1}]}^{(f\circ\phi_{\phi
\cdot\tilde{c},\tilde{c}})_{v,3}^{1,2}[N]}\right)  |\phi_{\tilde{c}_{A}%
,\phi\cdot\tilde{c}_{A}}\cdot c_{A}\rangle &  &
\end{array}
\end{equation}

Further analysis of above equation can be divided into following two (mutually
exclusive and exhaustive) cases exactly as in the previous section.\newline%
\begin{equation}%
\begin{array}
[c]{lll}%
\text{\underline{\textbf{Case 1}} :}\ \phi_{\tilde{c}_{A},\phi\cdot\tilde
{c}_{A}}\cdot c_{A}\ \notin\ \phi_{\tilde{c},\phi\cdot\tilde{c}}\cdot
\lbrack\tilde{c}_{1}] &  & \\
\vspace*{0.1in}\hspace*{0.5in}\Leftrightarrow\ \phi\cdot\tilde{c}_{A}%
\ \neq\ \phi\cdot\tilde{c}\ \Leftrightarrow\ \tilde{c}_{A}\ \neq\ \tilde{c} &
&
\end{array}
\end{equation}
In this case it is clear that LHS = 0.\newline%
\begin{equation}%
\begin{array}
[c]{lll}%
\text{\underline{\textbf{Case 2}} :}\ \phi_{\tilde{c}_{A},\phi\cdot\tilde
{c}_{A}}\cdot c_{A}\ \in\ \phi_{\tilde{c},\phi\cdot\tilde{c}}\cdot
\lbrack\tilde{c}_{1}] &  & \\
\vspace*{0.1in}\hspace*{0.5in}\Leftrightarrow\ \phi\cdot\tilde{c}_{A}%
\ =\ \phi\cdot\tilde{c}\ \Leftrightarrow\ \tilde{c}_{A}\ =\ \tilde{c} &  &
\end{array}
\end{equation}
In this case,
\begin{equation}%
\begin{array}
[c]{lll}%
\boxed{\textrm{LHS}\ =\ \sum_{v\in V(\tilde{c})}\left(\left(f\circ\phi_{\phi\cdot\tilde{c},\tilde{c}}\right)^{1,2}_{v,2}[N]\left(\overline{V}(\phi_{\tilde{c}_{A},\phi\cdot\tilde{c}_{A}})\right)\ -\ \left(f\circ\phi_{\phi\cdot\tilde{c},\tilde{c}}\right)^{1,2}_{v,3}[N]\left(\overline{V}(\phi_{\tilde{c}_{A},\phi\cdot\tilde{c}_{A}})\right)\right)} &
&
\end{array}
\end{equation}
For the sake of pedagogy, and without any loss in generality we again assume
(this assumption was also made in the previous section) that the only vertex
in $V(\phi\cdot\tilde{c})\ =\ V(\phi\cdot\tilde{c}_{A})$ which falls inside
the support of $N$ is a vertex $v_{A}$.\newline In this case \textbf{Case 2}
gets further sub-divided into following two complementary cases.\newline%
\underline{\textbf{case 2a}} : $\phi_{\tilde{c}_{A},\phi\tilde{c}_{A}}\cdot
c_{A}$ does not contain an EO vertex $(v_{A})_{\delta}^{E}$ of type-$(1,2)$ in
the neighbourhood of $v_{A}$.\newline In this case the vertex functions are
unchanged, and
\begin{equation}%
\begin{array}
[c]{lll}%
\text{LHS}\ =\ \left(  f\circ\phi_{\phi\cdot\tilde{c},\tilde{c}}\right)
_{v_{A},2}^{1,2}[N]\left(  \overline{V}(\phi_{\tilde{c}_{A},\phi\cdot\tilde
{c}_{A}})\right)  \ -\ \left(  f\circ\phi_{\phi\cdot\tilde{c},\tilde{c}%
}\right)  _{v_{A},3}^{1,2}[N]\left(  \overline{V}(\phi_{\tilde{c}_{A}%
,\phi\cdot\tilde{c}_{A}})\right)  &  & \\
\vspace*{0.1in}\hspace*{0.5in}=\ \left(  f\left(  \overline{V}(\phi_{\tilde
{c}_{A},\phi\cdot\tilde{c}_{A}})\right)  \ -\ f\left(  \overline{V}%
(\phi_{\tilde{c}_{A},\phi\cdot\tilde{c}_{A}})\right)  \right)  &  & \\
\vspace*{0.1in}\hspace*{0.5in}=\ 0 &  &
\end{array}
\end{equation}
\underline{\textbf{case 2b}} : $\phi_{\tilde{c}_{A},\phi\tilde{c}_{A}}\cdot
c_{A}$ contains an EO vertex $(v_{A})_{\delta}^{E}$ of type $(1,2)$ in the
neighbourhood of $v_{A}$ for some $\delta$.\newline In this case we can use
({\ref{h2psi1f1}}) and ({\ref{h2psi1f3}}) to get,
\begin{equation}%
\begin{array}
[c]{lll}%
\text{LHS}\ =\ \left(  f\circ\phi_{\phi\cdot\tilde{c},\tilde{c}}\right)
_{v_{A},2}^{1,2}[N]\left(  \overline{V}(\phi_{\tilde{c}_{A},\phi\cdot\tilde
{c}_{A}})\right)  \ -\ \left(  f\circ\phi_{\phi\cdot\tilde{c},\tilde{c}%
}\right)  _{v_{A},3}^{1,2}[N]\left(  \overline{V}(\phi_{\tilde{c}_{A}%
,\phi\cdot\tilde{c}_{A}})\right)  &  & \\
\vspace*{0.1in}\ =\left[  \lambda(\vec{n}_{\tilde{c}}^{v_{A}%
})\ V_{1}^{a}(v_{A},\phi\cdot\tilde{c})\ \frac{\partial}{\partial(x_{v_{A}%
}^{\phi\cdot\tilde{c}})^{a}}N(v_{A},\{x_{v_{A}}\}^{\phi\cdot\tilde{c}%
})\ \left(  f\circ\phi_{\phi\cdot\tilde{c},\tilde{c}}\right)  \left(
(v_{A})_{\delta}^{E}(1,3),\overline{V}(\phi_{\tilde{c}_{A},\phi\cdot
\tilde{c}_{A}})-\{(v_{A})_{\delta}^{E}(1,2)\}\right)  \right.  &  & \\
\hspace*{0.4in}-\left.  \lambda(\vec{n}_{\tilde{c}}^{v_{A}})\ V_{3}^{a}%
(v_{A},\phi\cdot\tilde{c})\nabla_{a}N(v_{A},\{x_{v_{A}}\}^{\phi\cdot\tilde{c}%
})\left(  f\circ\phi_{\phi\cdot\tilde{c},\tilde{c}}\right)  \left(
(v_{A})_{\delta}^{E}(1,1),\overline{V}(\phi_{\tilde{c}_{A},\phi\cdot
\tilde{c}_{A}})-\{(v_{A})_{\delta}^{E}(1,2)\}\right)  \right]  &  & \\
&  &
\end{array}
\end{equation}

\begin{equation}
\label{case2bLHSH2onS1}%
\begin{array}
[c]{lll}%
\text{LHS}=\\
\vspace*{0.1in}
\left[  \lambda(\vec{n}_{\tilde{c}}^{v_{A}})\ V_{1}^{a}%
(v_{A},\phi\cdot\tilde{c})\ \frac{\partial}{\partial(x_{v_{A}}^{\phi
\cdot\tilde{c}})^{a}}N(v_{A},\{x_{v_{A}}\}^{\phi\cdot\tilde{c}})\ \left(
f\circ\phi_{\phi\cdot\tilde{c},\tilde{c}}\right)  \left(  (v_{A})^{E}_{\delta
}(1,3),\overline{V}(\phi_{\tilde{c}_{A},\phi\cdot\tilde{c}_{A}}%
)-\{(v_{A})^{E}_{\delta}(1,2)\}\right)  \right.  &  & \\
\hspace*{0.4in}-\left.  \lambda(\vec{n}_{\tilde{c}}^{v_{A}})\ V_{3}^{a}%
(v_{A},\phi\cdot\tilde{c})\nabla_{a}N(v_{A},\{x_{v_{A}}\}^{\phi\cdot\tilde{c}%
})f\left(  (\phi^{-1}\cdot v_{A})^{E}_{\delta}(1,1),\overline{V}%
(c_{A})-\{(\phi^{-1}\cdot v_{A})^{E}_{\delta}(1,2)\}\right)  \right]  &
&
\end{array}
\end{equation}
where in the last line we have used the key property of our new
representation. If $(v_{A})^{E}_{\delta}(1,2)$ is an EO vertex in
$V(\phi_{\tilde{c},\phi\cdot\tilde{c}}\cdot c_{A})$ which is associated to
$v_{A}$ (which is in turn a vertex in $V(\phi\cdot\tilde{c}_{A})$ then,
\begin{equation}%
\begin{array}
[c]{lll}%
\phi_{\phi\cdot\tilde{c},\tilde{c}}\cdot(v_{A})^{E}_{\delta}%
(1,2)\ =\ (\phi^{-1}\cdot v_{A})^{E}_{\delta}(1,2) &  &
\end{array}
\end{equation}
that is, $\phi_{\phi\cdot\tilde{c},\tilde{c}}$ maps it to an EO vertex in
$V(\phi_{\phi\cdot\tilde{c},\tilde{c}})$ which is associated to $\phi
^{-1}\cdot v_{A}\ \in\ V(\tilde{c})$.\newline

We now analyze the RHS and show that in all the three cases ( (\textbf{case
1}), (\textbf{case 2a}), (\textbf{case 2b}) ), it matches the LHS answers
given above.\newline%
\begin{equation}%
\begin{array}
[c]{lll}%
\text{RHS}\ =\ \left(  \hat{H}^{2}[(\phi_{\tilde{c},\phi\cdot\tilde{c}}%
)^{*}N]\Psi^{f^{1}}_{[\tilde{c}]_{1}}\right)  \vert c_{A}\rangle &  &
\end{array}
\end{equation}
It is clear that in the first case, (\textbf{case 1}), as $\tilde{c}%
\ \neq\ \tilde{c}_{A}$ clearly
\begin{equation}
\nonumber\\
\text{RHS}\ =\ 0
\end{equation}
Now consider (\textbf{case 2a}).
\begin{equation}%
\begin{array}
[c]{lll}%
\text{\underline{\textbf{Case 2a}} :}\ \phi_{\tilde{c}_{A},\phi\cdot\tilde
{c}_{A}}\dot c_{A}\ \in\ \phi_{\tilde{c},\phi\cdot\tilde{c}}\cdot[\tilde
{c}_{1}] &  & \\
\vspace*{0.1in} \hspace*{0.5in} \Leftrightarrow\ \tilde{c}\ =\ \tilde{c}_{A} &
&
\end{array}
\end{equation}
But $\phi_{\tilde{c}_{A},\phi\tilde{c}_{A}}\cdot c_{A}$ does not contain an EO
vertex $(v_{A})^{E}_{\delta}$ of type-$(1,2)$ in the neighbourhood of $v_{A}$.
Where $v_{A}$ is the only vertex of $\phi\cdot\tilde{c}$ which lies inside the
support of $N$. Obviously this implies that the only vertex of $\tilde{c}$
which lies inside the support of $\phi_{\tilde{c},\phi\cdot\tilde{c}}^{*}N$ is
$\phi^{-1}\cdot v_{A}$.\newline Now notice that as as the EO structure
associated to any charge-network $c$ is preserved under the $\phi_{\tilde
{c},\phi\cdot\tilde{c}}$ for any diffeomorphism $\phi$ (as demonstrated in
equation (\ref{preserveEO-1})), $c_{A}$ does not contain an EO vertex of
type-$(1,2)$ in the nighrbourhood of $\phi^{-1}\cdot v_{A}$, whence in this
case
\begin{equation}%
\begin{array}
[c]{lll}%
\text{RHS}\ =\ f^{1,2}_{\phi^{-1}\cdot v_{A},2}[(\phi_{\tilde{c},\phi
\cdot\tilde{c}})^{*}N]\left(  \overline{V}(c_{A})\right)  \ -\ f^{1,2}%
_{\phi^{-1}\cdot v_{A},3}[(\phi_{\tilde{c},\phi\cdot\tilde{c}})^{*}N]\left(
\overline{V}(c_{A})\right)  &  & \\
\vspace*{0.1in} \hspace*{0.5in}=\ 0 &  &
\end{array}
\end{equation}
Recall that even the LHS was trivial in this case.\newline We now turn to the
remaining case, (\textbf{case 2b}) for which LHS was non-trivial. For the
benefit of reader, we recall the conditions defining this case again.\newline%
\underline{\textbf{case 2b}} : $\phi_{\tilde{c}_{A},\phi\tilde{c}_{A}}\cdot
c_{A}$ contains an EO vertex $(v_{A})^{E}_{\delta}$ of type-$(1,2)$ in the
neighbourhood of $v_{A}$ for some $\delta$.\newline Once again, using equation
(\ref{preserveEO-1}) we see that $c_{A}$ contains an EO vertex $(\phi
^{-1}\cdot v_{A})^{E}_{\delta}$ of type-$(1,2)$ in the neighborhood of
$\phi^{-1}\cdot v_{A}$ \emph{for the same $\delta$}. Hence in this case, RHS
is given by,
\begin{equation}
\label{case2bRHSH2onLMI1}%
\begin{array}
[c]{lll}%
\text{RHS}\ =\ f^{1,2}_{\phi^{-1}\cdot v_{A},2}[(\phi_{\tilde{c},\phi
\cdot\tilde{c}})^{*}N]\left(  \overline{V}(c_{A})\right)  \ -\ f^{1,2}%
_{\phi^{-1}\cdot v_{A},3}[(\phi_{\tilde{c},\phi\cdot\tilde{c}})^{*}N]\left(
\overline{V}(c_{A})\right)  &  & \\
\vspace*{0.1in} =\left[  \lambda(\vec{n}_{\tilde{c}}^{\phi^{-1}\cdot v_{A}%
})\ V_{1}^{a}(\phi^{-1}\cdot v_{A},\tilde{c})(\frac{\partial}{\partial
(x_{\phi^{-1}\cdot v_{A}}^{\tilde{c}})^{a}})\phi_{\tilde{c},\phi\cdot\tilde
{c}})^{*}N(\phi^{-1}\cdot v_{A},\{x_{\phi^{-1}\cdot v_{A}}\}^{\tilde{c}%
})\right.  &  & \\
\hspace*{1.5in}f^{1}\left(  (\phi^{-1}\cdot v_{A})^{E}_{\delta}%
(1,3),\overline{V}(c_{A})-\{(\phi^{-1}\cdot v_{A})^{E}_{\delta
}(1,2)\}\right)  &  & \\
-\lambda(\vec{n}_{\tilde{c}}^{\phi^{-1}\cdot v_{A}})\ V_{3}^{a}(\phi^{-1}\cdot
v_{A},\tilde{c})(\frac{\partial}{\partial(x_{\phi^{-1}\cdot v_{A}}^{\tilde{c}%
})^{a}})(\phi_{\tilde{c},\phi\cdot\tilde{c}})^{*}N(\phi^{-1}\cdot
v_{A},\{x_{\phi^{-1}\cdot v_{A}}\}^{\tilde{c}}) &  & \\
\hspace*{1.5in} \left.  f^{1}\left(  (\phi^{-1}\cdot v_{A})^{E}_{\delta
}(1,1),\overline{V}(c_{A})-\{(\phi^{-1}\cdot v_{A})^{E}_{\delta
}(1,2)\}\right)  \right]  &  &
\end{array}
\end{equation}
We can use
\begin{equation}%
\begin{array}
[c]{lll}%
\lambda(\vec{n}_{\tilde{c}}^{\phi^{-1}\cdot v_{A}})\ =\ \lambda(\vec{n}%
_{\phi\cdot\tilde{c}}^{v_{A}}) &  & \\
\vspace*{0.1in} V^{a}_{i}(\phi^{-1}\cdot v_{A},\tilde{c})\ =\ V^{a}_{i}%
(v_{A},\phi\cdot\tilde{c})\ \forall\ a,\ i &  & \\
\vspace*{0.1in} \frac{\partial}{\partial(x_{\phi^{-1}\cdot v_{A}}^{\tilde{c}%
})^{a}}\ =\ (\phi_{\phi\cdot\tilde{c},\tilde{c}})_{*}\frac{\partial}%
{\partial(x_{v_{A}}^{\phi\cdot\tilde{c}})^{a}} &  &
\end{array}
\end{equation}
to simplify (\ref{case2bRHSH2onLMI1})
\begin{equation}
\label{case2bRHSH2onLMI1-2}%
\begin{array}
[c]{lll}%
\text{RHS}\ =\ \left[  \lambda(\vec{n}_{\phi\cdot\tilde{c}}^{v_{A}}%
)\ V_{1}^{a}(v_{A},\phi\cdot\tilde{c})\left(  \frac{\partial}{\partial
(x_{v_{A}}^{\phi\cdot\tilde{c}})^{a}}\right)  (\phi_{\phi\cdot\tilde{c}%
,\tilde{c}}^{*}\phi_{\tilde{c},\phi\cdot\tilde{c}})^{*}N(v_{A},\{x_{v_{A}%
}\}^{\phi\cdot\tilde{c}})\right.  &  & \\
\hspace*{1.5in}f^{1}\left(  (\phi^{-1}\cdot v_{A})^{E}_{\delta}%
(1,3),\overline{V}(c_{A})-\{(\phi^{-1}\cdot v_{A})^{E}_{\delta}(1,
2)\}\right)  &  & \\
\hspace*{0.6in}-\lambda(\vec{n}_{\phi\cdot\tilde{c}}^{v_{A}})\ V_{3}^{a}%
(v_{A},\phi\cdot\tilde{c})\left(  \frac{\partial}{\partial(x_{v_{A}}%
^{\phi\cdot\tilde{c}})^{a}}\right)  (\phi_{\phi\cdot\tilde{c},\tilde{c}}%
^{*}\phi_{\tilde{c},\phi\cdot\tilde{c}})^{*}N(v_{A},\{x_{v_{A}}\}^{\phi
\cdot\tilde{c}}) &  & \\
\hspace*{1.5in}\left.  f^{1}\left(  (\phi^{-1}\cdot v_{A})^{E}_{\delta
}(1,1),\overline{V}(c_{A})-\{(\phi^{-1}\cdot v_{A})^{E}_{\delta}(1,
2)\}\right)  \right]  &  & \\
\vspace*{0.2in} \text{RHS}=\ \lambda(\vec{n}_{\phi\cdot\tilde{c}}^{v_{A}%
})\left[  \ V_{1}^{a}(v_{A},\phi\cdot\tilde{c})\left(  \frac{\partial
}{\partial(x_{v_{A}}^{\phi\cdot\tilde{c}})^{a}}\right)  N(v_{A},\{x_{v_{A}%
}\}^{\phi\cdot\tilde{c}})\right.  &  & \\
\hspace*{1.5in}f^{1}\left(  (\phi^{-1}\cdot v_{A})^{E}_{\delta}%
(1,3),\overline{V}(c_{A})-\{(\phi^{-1}\cdot v_{A})^{E}_{\delta}(1,
2)\}\right)  &  & \\
\hspace*{0.6in}-\ V_{3}^{a}(v_{A},\phi\cdot\tilde{c})\left(  \frac{\partial
}{\partial(x_{v_{A}}^{\phi\cdot\tilde{c}})^{a}}\right)  N(v_{A},\{x_{v_{A}%
}\}^{\phi\cdot\tilde{c}}) &  & \\
\hspace*{1.5in}\left.  f^{1}\left(  (\phi^{-1}\cdot v_{A})^{E}_{\delta
}(1,1),\overline{V}(c_{A})-\{(\phi^{-1}\cdot v_{A})^{E}_{\delta}(1,
2)\}\right)  \right]  &  &
\end{array}
\end{equation}
On comparing (\ref{case2bRHSH2onLMI1-2}) with (\ref{case2bLHSH2onS1}) we
conclude that even in this case (\textbf{case 2b})
\begin{equation}
\nonumber\\
\text{LHS}\ =\ \text{RHS}%
\end{equation}
Whence, we finally have%

\begin{equation}
\label{covofH2onS1-final}\boxed{\hat{U}(\phi)^{\prime}\hat{H}^{(2)}[N]^{\prime}\hat{U}(\phi^{-1})^{\prime}\Psi_{\lbrack\tilde{c}^{1}]_{1}}^{f^{1}} =\ \hat{H}^{(2)}[(\phi_{\tilde{c},\phi\cdot\tilde{c}})^{\ast}N]^{\prime}\Psi _{\lbrack\tilde{c}^{1}]_{1}}^{f^{1}}}
\end{equation}
$\forall\ \phi$.

One can similarly show that%

\begin{equation}
\label{covofH3onS1-final}\boxed{\hat{U}(\phi)^{\prime}\hat{H}^{(3)}[N]^{\prime}\hat{U}(\phi^{-1})^{\prime}\Psi_{\lbrack\tilde{c}^{1}]_{1}}^{f^{1}} =\ \hat{H}^{(3)}[(\phi_{\tilde{c},\phi\cdot\tilde{c}})^{\ast}N]^{\prime}\Psi _{\lbrack\tilde{c}^{1}]_{1}}^{f^{1}}}
\end{equation}
$\forall\ \phi$.

Using (\ref{covofH1onS1-final}), (\ref{covofH2onS1-final}) and
(\ref{covofH3onS1-final}) we see that%

\begin{equation}
\label{diffcovofHonS1}\hat{U}(\phi)^{\prime}\hat{H}[N]^{\prime}\hat{U}%
(\phi^{-1})^{\prime}\Psi_{\lbrack\tilde{c}^{1}]_{1}}^{f^{1}} =\ \hat{H}%
[(\phi_{\tilde{c},\phi\cdot\tilde{c}})^{\ast}N]^{\prime}\Psi_{\lbrack\tilde
{c}^{1}]_{1}}^{f^{1}}%
\end{equation}
$\forall\ \phi$.

It is straightforward to generalize this result to $\Psi^{f^{i}}_{[\tilde
{c}]_{i}}\ \forall\ i$.\newline

Above result, in conjunction with (\ref{indeedarep}) and result of \cite{hat}
shows that we have a representation of Dirac algebra on $\mathcal{V}_{LMI}$ in
the loop quantized $2+1$ dimensional $U(1)^{3}$ theory.\newline

\section{Spectrum of the theory}\label{spectrum}

The new representation of $Diff(\Sigma)$ on ${\cal V}_{LMI}$ which was a crucial ingredient in establishing the diffeomorphism covariance of $H[N]$ required us to introduce certain auxiliary structures. The choice of reference charge-networks for each gauge orbit $[\tilde{c}]_{diff}$ 


We analyze the spectrum of the theory in order to probe the viability of new representation for $Diff(\Sigma)$. Let us try to find a simplest class of states in ${\cal V}_{LMI}$ which are solutions to $\hat{H}[N]$. Consider a class of states of the form
\begin{equation}\label{classofstates}
\begin{array}{lll}
\vert\Phi\rangle\ =\ \sum_{I=1}^{N} \sum_{m=1}^{3}\left[a^{(m)}_{I}\ \Psi^{(f_{I})^{m}}_{[\tilde{c}_{I}]_{m}}\right]
\end{array}
\end{equation}
where $1\ \leq N\ <\ \infty$. This is a fairly large class of states in which we look for states which satisfy
\begin{equation}
\sum_{i=1}^{3}\hat{H}^{i}[N]\vert\Phi\rangle\ =\ 0
\end{equation}
$\forall\ N$.
The resulting equation can be written  in a condensed form as,
\begin{equation}\label{keyeq}
\begin{array}{lll}
\sum_{i=1}^{3}\hat{H}^{i}[N]\vert\Phi\rangle\ =\\
\vspace*{0.1in}
\hspace*{0.8in}\sum_{I=1}^{N}\sum_{v\in V(\tilde{c}_{I})}\sum_{i,j=1}^{3}a_{I}^{i}\left[\epsilon_{ijk}\Psi^{(f_{I})^{i,i}_{v,j}[N]}_{[\tilde{c}_{I}]_{i}}\ +\ \epsilon_{ijk}\left(\Psi^{(f_{I})^{i,j}_{v,k}[N]}_{[\tilde{c}_{I}]_{i}}\ -\ \Psi^{(f_{I})^{i,j}_{v,i}[N]}_{[\tilde{c}_{I}]_{i}}\right)\right]
\end{array}
\end{equation}
From here it is easy to see that, in the class of states given in (\ref{classofstates})  there is a subset obtained by choosing $f_{I}\ =\ \textrm{constant}\ \textrm{and}\ a_{I}^{1}\ =\ a_{I}^{2}\ =\ a_{I}^{3}\ \forall\ I$ which lie in the kernel of the Hamiltonian constraint. This result is not completely expected a priori as action of $H^{j}[N]$ on $\Psi^{f^{k}}_{[\tilde{c}]_{k}}$ is not trivial even when the vertex function $f^{k}$ is taken to be constant when $j\ \neq\ k$. The anti-symmetry in the internal indices in the Hamiltonian constraint (which is rather neatly encoded in this expression) is responsible for the fact that $\sum_{i=1}^{3}\Psi_{[\tilde{c}]_{i}}^{f^{i}=\textrm{const}}$ lie in the kernel of Hamiltonian constraint.\\
As the set $[\tilde{c}]_{i}$ is diffeomorphism (intact homeomorphism) invariant, we can see that (formally)  identifying all the Habitat states which are related by diffeomorphisms will yield distributions on ${\cal H}_{kin}$ of the type
\begin{equation}
\begin{array}{lll}
\vert[\Phi]\rangle\ =\ \sum_{i=1}^{3}\sum_{[\tilde{c}^{\prime}]_{i}\vert\tilde{c}^{\prime}\ =\ \phi\cdot\tilde{c}}\Psi_{[\tilde{c}^{\prime}]_{i}}^{f^{i}=\textrm{const}}\\
\vspace*{0.1in}
\hspace*{0.5in}=\ \sum_{i=1}^{3}\sum_{\tilde{c}^{\prime}\ =\ \phi\cdot\tilde{c}}\Psi_{[\tilde{c}^{\prime}]_{i}}^{f^{i}=\textrm{const}}\\
\vspace*{0.1in}
\hspace*{0.1in}=\ \sum_{i=1}^{3}\sum_{\tilde{c}^{\prime}\ =\ \phi_{\tilde{c},\phi\cdot\tilde{c}}\cdot\tilde{c}}\Psi_{[\tilde{c}^{\prime}]_{i}}^{f^{i}=\textrm{const}}
\end{array}
\end{equation}
Whence the sum over diffeomorphisms reduces to summing over the $\phi-maps$ as each $[\tilde{c}]_{i}$ has a unique vertex-free state $\tilde{c}$ associated to it. We see this as a hint that as far as the spectrum of the theory is concerned, summing over all diffeomorphisms might be equivalent to summing over the selected set of diffeomorphisms as dictated by the new representation.\\

\section{Conclusion and outlook}

In this paper, we continued our construction of a representation of Dirac algebra in quantum $U(1)^{3}$ gauge theory which was initiated in \cite{hat}. We considered the Hamiltonian constraint $\hat{H}[N]$  defined in \cite{hat} and constructed a representation of the Diffeomorphism group on the LMI-habitat ${\cal V}_{LMI}$ such that $\hat{U}(\phi), \hat{H}[N]$ satisfy the off-shell closure condition. In contrast to the original Hamiltonian constraint of Thiemann (constructed remarkably for the case of four dimensional LQG) where diffeomorphism covariance followed as a result of,  {\bf (i)} A diffeomorphism covariant choice of (state dependent) triangulation, {\bf (ii)} By assigning state dependent neighborhoods to each vertex of the spin-network such that this assignment was diffeomorphism invariant ; we had to introduce several new ingredients not least  of which is a new representation of the Diffeomorphism group. From the point of view of ${\cal H}_{kin}$ this representation is completely ad-hoc (and not even unitary !),  however this is no longer relevant to us as in our scheme, the ``kinematical" arena (the space on which quantum constraints are defined) is played out by ${\cal V}_{LMI}$. It is here where this representation is unitary with respect to the canonical inner product.\\
In addition to the new representation for $Diff(\Sigma)$ we also introduced a notion of state-dependent atlas on $\Sigma$. Roughly speaking the idea is to fix an atlas ${\cal U}(\Sigma, \tilde{c}_{0})$ for each reference charge-net $\tilde{c}_{0}$ ( one reference charge network associated to  each diffeomorphism-invariant orbit $[\tilde{c}]_{diff}$ of WEO vertex-free charge-networks.) and then for any $\tilde{c}^{\prime}\ \in\ [\tilde{c}]_{diff}$ we defined an atlas ${\cal U}(\Sigma, \tilde{c}^{\prime})$ associated to $\tilde{c}^{\prime}$ by pushing forward ${\cal U}(\Sigma, \tilde{c}_{0})$ using $\phi_{\tilde{c}_{0},\tilde{c}^{\prime}}$. The new representation together with the state dependence of co-ordinate charts ensured that extra-ordinariness of a vertex is an diffeomorphism invariant notion. This was crucial in establishing diffeomorphism covariance of the Hamiltonian constraint.

The use of new representation of the diffeomorphism group may seem worrisome as the canonical representation used so far in LQG has been analyzed in great detail and whose solution lead to generalized knot classes. In order to analyze the validity of the new representation we considered solving the Hamiltonian constraint in ${\cal V}_{LMI}$ and ask if the states obtained by ``formally" averaging over all diffeomorphisms would agree with states obtained by averaging over the preferred set. As we saw, for the $2+1$ dimensional theory, these results do in fact match for a subspace of kernel that we computed in section (\ref{spectrum}). This merely represents a small check on the validity of the new representation of $DIff(\Sigma)$ on ${\cal V}_{LMI}$. The issue however needs further investigation. 
As we have seen, the requirement that Hamiltonian constraint be diffeomorphism covariant on ${\cal V}_{LMI}$ is quite a stringent requirement and certainly reduces the vast amount of ambiguity which was present in quantization of $\hat{H}[N]$ presented in \cite{hat}. The main source of ambiguity in the definition of $\hat{H}[N]$ was in the determination of quantum shift vectors. 
As the definition of quantum shift is regularization dependent, in principle one can associate to each WEO-vertex free charge network $\tilde{c}$ a different regularization scheme for computing the quantum shift. However as we saw above, diffeomorphism covariance of $\hat{H}[N]$ requires determination of quantum shift \emph{only} on reference charge-nets in each diffeomorphism invariant orbit. For any other charge-net the quantum shift vectors are uniquely determined via push-forwards.\\
Perhaps the most un-satisfactory part of our construction is that our final definition of quantum constraints (or finite transformations generated by them) depends on various auxiliary structures. We list them below.\\
\noindent{(1)} The choice of reference charge-network $\tilde{c}_{0}$ in each orbit of WEO-vertex free charge nets.\\
\noindent {(2)} The choice of the subcategory in $[\tilde{c}]_{diff}$ or equivalently choice of set of diffemorphisms $\phi_{\tilde{c}_{0},\tilde{c}^{\prime}}$ associated to each diffeomorphism invariant orbit.\\
\noindent {(3)} The choice of co-ordinate atlas ${\cal U}(\Sigma, \tilde{c}_{0})$ for each $\tilde{c}_{0}$.\\
These structures can also be thought of as the data parametrizing quantization ambiguities which are input in the definition of quantum constraints.\\
A key open question is if the use of this auxiliary structures is viable. The final answer to this question can only be obtained by looking at expectation value of observables in the physical Hilbert space which should not depend on any ad-hoc intermediate structures.\\
We believe that the work we have done here admits a possibility of generalization to Euclidean Quantum gravity. An extremely important aspect to keep in mind here is that the geometric action of Hamiltonian constraint in $SU(2)$ case can also be understood in terms of phase-space dependent diffeomorphism on the dynamical fields \cite{abhay}. In light of this  result one could seek a quantization of Hamiltonian constraint in $SU(2)$ theory with the key lesson being provided by equation (\ref{keyeq}). The structure of internal indices show a tempting possibility of how the extension to $SU(2)$ may be possible.\\ 
In any event we believe that some of the lessons we have learnt here as well as in \cite{hat} together with the lessons learnt in (\cite{madtom}, \cite{mad2}) will have implications in defining quantum dynamics in  canonical Loop Quantum Gravity.

\section*{Acknowledgement}
We are indebted to Madahavan Varadarajan for sharing his results in \cite{mad2} with us prior to publication and also explaining to us the use of state-dependent coordinate charts as a tool in constructing finite triangulation Hamiltonian constraint. AL is also indebted to Abhay Ashtekar for sharing his insights regarding $SU(2)$ theory and his constant encouragement. We thank Miguel Campiglia for many discussions  over the course of this work and support. Work of AL is supported by Ramanujan fellowship of the Department of Science and Technology. Work of CT is supported by NSF grant PHY-0748336, a Penn State Mebus Fellowship, and a DAAD research grant.


%






\section*{Appendix}

\appendix
In this Appendix we prove a lemma which is used crucially in section (\ref{H1onf1}).\\

\textbf{Lemma A.1}\label{uniquenessoftildestate} : Let $\tilde{c}_{A}$, $\tilde{c}$
be WEO-vertex free states and let $c_{A}\ \in\ [\tilde{c}_{A}]_{i}%
,\ c\ \in\ [\tilde{c}]_{i}$. If for any diffeomorphism $\phi$
\begin{equation}
\phi_{\tilde{c}_{A},\phi\cdot\tilde{c}_{A}}\cdot c_{A}\ \in\ \phi_{\tilde
{c},\phi\cdot\tilde{c}}\cdot\lbrack\tilde{c}]_{i}%
\end{equation}
then $\tilde{c}_{A}\ =\ \tilde{c}$. \newline\textbf{Proof} : \newline Without
loss of generality we assume that $i=1$. We will also assume that $\tilde{c}$
has only one WE vertex. That is, $\exists$ a $v_{0}\ \in V(\tilde{c})$ such
that $(v_{0},v_{0}^{\prime})$ is the WEO pair in $c$ with $v_{0}^{\prime}$
being WE vertex of type-1.\newline We also recall some notations from Section
\ref{RECAP}. \noindent{(1)} As $c_{A}\ \in\ [\tilde{c}_{A}]_{1}$
\begin{equation}
c_{A}\ =\ (c_{A1},\tilde{c}_{A2},\tilde{c}_{A3})
\end{equation}
\noindent{(2)} Any $c^{\prime}\ \in\ \phi_{\tilde{c},\phi\cdot\tilde{c}}%
\cdot\lbrack\tilde{c}]_{1}$ is of the form
\begin{equation}
c^{\prime}\ =\ (\phi_{\tilde{c},\phi\cdot\tilde{c}}\cdot c_{1}^{\prime}%
,\phi\cdot\tilde{c}_{2},\phi\cdot\tilde{c}_{3})
\end{equation}

Hence we have%

\begin{equation}
\tilde{c}_{A i}\ =\ \tilde{c}_{i}\ \text{for}\ i=2,3
\end{equation}

Thus we have the following
\begin{equation}%
\begin{array}
[c]{lll}%
\phi_{\tilde{c}_{A},\phi\cdot\tilde{c}_{A}}(c_{A 1},\tilde{c}_{2},\tilde
{c}_{3})\ =\ \phi_{\tilde{c},\phi\cdot\tilde{c}}(c_{1},\tilde{c}_{2},\tilde
{c}_{3}) &  & \\
\vspace*{0.1in}(\tilde{c}_{A1},\tilde{c}_{2},\tilde{c}_{3})\ \in\lbrack
\tilde{c}]_{diff},\ \tilde{c}_{A1}\ \neq\ \tilde{c}_{1} &  &
\end{array}
\end{equation}
As $\gamma(\tilde{c}_{A1}\cup\tilde{c}_{2}\cup\tilde{c}_{3})\ \subset
\ \gamma(c_{A1}\cup\tilde{c}_{2}\cup\tilde{c}_{3})$, $\exists$ a
$(\overline{c}_{1},\tilde{c}_{2},\tilde{c}_{3})|\ \gamma(\overline{c}_{1}%
\cup\tilde{c}_{2}\cup\tilde{c}_{3})|\subset\ \gamma(c_{1}\cup\tilde{c}_{2}%
\cup\tilde{c}_{3})$ such that
\begin{equation}%
\begin{array}
[c]{lll}%
\phi_{\tilde{c}_{A},\phi\cdot\tilde{c}_{A}}(\tilde{c} _{A1},c_{A2}%
,c_{A3})\ =\ \phi\cdot(\tilde{c}_{1},\tilde{c}_{2},\tilde{c}_{3}%
)\ =\ \phi_{\tilde{c},\phi\cdot\tilde{c}}(\overline{c}_{1},\tilde{c}%
_{2},\tilde{c}_{3}) &  &
\end{array}
\end{equation}
where (as a trivial consequence of above equation) we have, \noindent{(1)}
$(\overline{c}_{1},\tilde{c}_{2},\tilde{c}_{3})$ is gauge-invariant.\newline%
\noindent{(2)} $(\overline{c}_{1},\tilde{c}_{2},\tilde{c}_{3})$ has no WE
vertex.\newline But from the above lemma we know that there is a unique
charge-network contained associated to $(c_{1},\tilde{c}_{2},\tilde{c}_{3})$
which satisfies above two conditions, and that is $\tilde{c}$. Whence we
have,
\begin{equation}%
\begin{array}
[c]{lll}%
\phi_{\tilde{c}_{A},\phi\cdot\tilde{c}_{A}}(\tilde{c}_{A 1} ,\tilde{c}%
_{2},\tilde{c}_{3})\ =\ \phi\cdot(\tilde{c}_{A 1},\tilde{c}_{2},\tilde{c}%
_{3})= &  & \\
\vspace*{0.1in}\hspace*{1in}\phi_{\tilde{c},\phi\cdot\tilde{c}}(\overline
{c}_{1},\tilde{c}_{2},\tilde{c}_{3})\ =\ \phi_{\tilde{c},\phi\cdot\tilde{c}%
}\cdot\tilde{c}\ =\ \phi\cdot\tilde{c} &  &
\end{array}
\end{equation}
Hence
\begin{equation}
\tilde{c}_{A}\ =\ \tilde{c}%
\end{equation}
q.e.d \newline\newline\newline

\end{document}